\pdfoutput=1
\documentclass
[aps,prd,nofootinbib,
preprint,
superscriptaddress,preprintnumbers,
]{revtex4}
\usepackage{graphicx,subfigure,latexsym}
\usepackage{amsmath,amssymb,amsfonts}
\usepackage{bm}
\usepackage{hyperref}
\usepackage{ulem}
\usepackage{color}

\newcommand\redout{\bgroup\markoverwith
{\textcolor{red}{\rule[.5ex]{2pt}{0.4pt}}}\ULon}

\newcommand{\be}{\begin{equation}}
\newcommand{\ee}{\end{equation}}
\newcommand{\bear}{\begin{eqnarray}}
\newcommand{\eear}{\end{eqnarray}}
\newcommand{\ba}{\begin{array}}
\newcommand{\ea}{\end{array}}

 \newcommand{\OO}{{\cal O}}

\def \be {\begin{equation}}
\def \ee {\end{equation}}

\def \bes {\begin{subequations}}
\def \ees {\end{subequations}}

\def \pd {\partial}

\def \eq {Eq.~}

\def \cmw {v_{B}}
\def \<{\langle}
\def \>{\rangle}
\def \+{\dagger}
\def \({\left(}
\def \){\right)}
\def \[{\left[}
\def \]{\right]}

\def \a {\alpha}
\def \b {\beta}

\def \e {\epsilon}

\def \o {\omega}
\def \s {\sigma}

\def \vp {\bm{p}}

\def \vx {\bm{x}}

\def \vv {\bm{v}}

\def \vb {\bm{b}}
\def \vk {\bm{k}}

\def \vB {\bm{B}}
\def \vE {\bm{E}}

\def \hp {\hat{\vp}}

\def \anom {\text{anom}}
\def \norm {\text{norm}}

\def \eq {Eq.~}

\begin{document}

\preprint{RBRC-1106}

\title{Collective Modes of Chiral Kinetic Theory in Magnetic Field}
\author{Misha~Stephanov}
\affiliation{Department of Physics, University of Illinois, Chicago,
  Illinois 60607}
\author{Ho-Ung Yee}
\affiliation{Department of Physics, University of Illinois, Chicago,
  Illinois 60607}
\affiliation{RIKEN-BNL Research Center, Brookhaven National Laboratory
Upton, New York
11973-5000}
\author{Yi Yin}
\affiliation{Department of Physics,
Brookhaven National Laboratory, Upton, New York 11973-5000}

\date{December 2014}

\begin{abstract}

  We study collective excitations in systems described by chiral
  kinetic theory in external magnetic field. We consider
  high-temperature weak-coupling plasma, as well as high-density
  Landau Fermi liquid with interaction not restricted to be weak. We
  show that chiral magnetic wave (CMW) emerges in hydrodynamic regime
  (at frequencies smaller than collision relaxation rate) and the CMW
  velocity is determined by thermodynamic properties only. We find
  that in a plasma of opposite chiralities, at frequencies smaller
  than the chirality-flipping rate, the CMW excitation turns into a
  vector-like diffusion mode. In the interacting Fermi liquid, the CMW
  turns into the Landau zero sound mode in the high-frequency
  collisionless regime.

\end{abstract}

\maketitle

\section{Introduction}
\label{sec:intro}
Parity-violating transport  in the plasma of chiral fermions
originating from chiral anomaly and driven by magnetic field
attracted significant amount of interest recently.
Such phenomena
can be realized, for example, in the early stages of heavy-ion collisions at RHIC and LHC 
where a very strong magnetic field is created by large
ultrarelativistic nuclei \cite{Kharzeev:2007jp}. 
One such effect which may have a significant phenomenological importance is the Chiral Magnetic Effect (CME) 
\cite{Kharzeev:2007jp,Fukushima:2008xe,Son:2004tq,Son:2007ny,Vilenkin:1980fu,Kharzeev:2010gr}. The CME
causes vector (axial) current along the direction of the magnetic
field in the presence of axial (vector) chemical potential. In
heavy ion collisions this should lead to dipolar charge separation perpendicular to the reaction plane,
whose experimental signature may be observed in two particle azimuthal correlations of charged particles \cite{Voloshin:2004vk,Abelev:2009ac,Selyuzhenkov:2011xq,Adamczyk:2014mzf}. The CME has been demonstrated in various theoretical frameworks, such as in hydrodynamics \cite{Son:2009tf}, in lattice QCD \cite{Buividovich:2009wi,Abramczyk:2009gb,Yamamoto:2011gk,Buividovich:2013hza,Bali:2014vja}, in effective theories \cite{Sadofyev:2010is,Nair:2011mk}
and in the AdS/CFT correspondence \cite{Yee:2009vw,Rebhan:2009vc,Gorsky:2010xu,Hoyos:2011us}.

One important consequence of the CME 
is the existence of a
new type of collective propagating excitations of chiral charges:
the Chiral Magnetic Wave (CMW) \cite{Kharzeev:2010gd,Newman:2005hd}.
In heavy-ion collisions,
the CMW leads to a buildup of net electric quadrupole moment in the plasma fireball
\cite{Burnier:2011bf,Gorbar:2011ya,Yee:2013cya},
resulting in a charge-dependent elliptic flow of pions
\cite{Burnier:2011bf,Burnier:2012ae,Yee:2013cya}.
There are experimental results which appear to be in line with this picture
\cite{Wang:2012qs,Ke:2012qb,Shou:2014cua,Voloshin:2014gja}.

The aim of this paper is to study these effects in a microscopic framework to gain further insight on the physics of chiral anomaly
in high temperature and/or density regime.
Although in the low-frequency regime the effects of the chiral anomaly,
due to their topological nature, should not depend on the microscopic
nature of the interactions, 
these effects in high-frequency regime depend on the microscopic dynamics.
In this work, we study chiral transport phenomena in a weakly-coupled regime using kinetic theory.

There has been recent progress on the kinetic theory with massless chiral fermions
\cite{Son:2012wh,Stephanov:2012ki,Son:2012zy,Loganayagam:2012pz,Gao:2012ix,Chen:2014cla},
and in particular,
the non-equilibrium kinetic equation describing the motion of chiral particles
was derived.
A crucial ingredient in the chiral kinetic theory is the 
Berry's phase in momentum space which originates from the underlying
chiral spinor wave function of quasi-particles
\cite{Son:2012wh,Stephanov:2012ki}. Remarkably, the effects of quantum
chiral anomaly are accounted by the Berry's phase in such a  (semi-)classical  kinetic theory.
Of particular importance is the modification of the velocity (i.e., its
relation to the momentum) of a chiral particle in the presence of magnetic field due to the Berry's phase.
Such anomalous velocity will indeed play a central role in this paper.

We begin by studying the collective waves in a plasma of a single Weyl
fermion species in the regime of high temperature (and low chemical potential) in the presence of a magnetic field.
As the collective excitations such as the CMW are hydrodynamic modes,
we need to introduce a relaxation dynamics provided by the collision term in the Boltzmann equation, to be able to derive such modes in the kinetic theory.
We show that the group velocity of the chiral magnetic wave at long wavelengths is given universally by the magnetic field and charge susceptibility, without depending on explicit details of the collision term.
On the other hand,
kinetic theory as a microscopic theory allows us to study the dispersion relation beyond hydrodynamic regime, and
for this purpose, we
use the relaxation time approximation (RTA) as a simplest example for the collision term, which enables us to find some analytic solutions. 

QCD and similar vector-like theories contain chiral fermions (quarks)
with opposite chiralities -- a Dirac fermion consists of a left-handed
and a right-handed Weyl fermion. In non-abelian gauge theories, there
are topological sphaleron fluctuations that induce transitions between
opposite chirality massless fermions. These non-perturbative effects,
in addition to small quark masses, provide the dynamics of axial
charge relaxation and we will refer to these generically as
inter-chiral transitions.  

We study how the inter-chiral transitions
affect the collective modes of a high temperature plasma in a magnetic
field which contains both left- and right-handed chiral fermions. At
sufficiently low frequency (wave number) where the inter-chiral transitions
play a significant role, we find an interesting diffusive mode whose
diffusion coefficient is governed by dynamics of chiral anomaly. With
increasing wave number, we observe a transition of this mode into the
chiral magnetic wave.  A similar result was also found recently in
a holographic model study~\cite{Jimenez-Alba:2014iia}.

We also explore the collective modes in the opposite regime of a cold dense Fermi liquid of chiral fermions~\cite{Son:2012wh,Son:2012zy},
which could be of relevance to dense quark matter in compact stars.
In normal Fermi liquids, it is well-known that in the collisionless
high-frequency regime there exist collective propagating modes of
Fermi surface fluctuations -- the Landau's zero sound. 
The zero sound in a Fermi liquid of chiral fermions in the presence of
magnetic field was first studied in Ref.\cite{Gorsky:2012gi}. However,
only the collisionless regime neglecting collision terms in the
Boltzmann equation was considered. Since the CMW is a collective
hydrodynamic mode which can be seen only with a collision term, we
extend the analysis of Ref.\cite{Gorsky:2012gi} by including a collision term in the relaxation time approximation.
As a result we find an interesting transition of the chiral
magnetic wave in the hydrodynamic regime into the zero sound in the
collisionless regime as one increases the wave number.

The paper is organized as follows. In the next section, we give a brief review of the 
chiral kinetic theory recently developed in Ref.~\cite{Son:2012wh,Stephanov:2012ki} that most of our analysis is based on. In the subsequence section~\ref{sec:left},
we study the collective modes with magnetic field in the hot plasma of a single right-handed Weyl fermion species.
In section~\ref{sec:sphaleron},
we study the plasma of chiral fermions with opposite chiralities,
including inter-chiral (sphaleron) transitions
in the collision term.
In section~\ref{sec:fermi},
the collective waves with magnetic field in a chiral cold dense Fermi liquid are studied.
We conclude in section~\ref{sec:conclusion}.

\section{ Review of Chiral kinetic theory}
\label{sec:review}

This section gives a brief review on the chiral kinetic theory developed in Ref.~\cite{Son:2012wh,Stephanov:2012ki},
which also serves to set up the notations and conventions we use throughout the paper.

In general,
kinetic theory describes the motion of weakly interacting quasi-particles in the regime where collisions are rare enough
so that the average distance traveled by quasi-particles between each collision is larger
than the quantum wavelength of the particles,
and the motions between collisions can be treated classically. 
This allows one to introduce a distribution function of quasi-particles in $(\vx,\vp)$ phase space,
and the dynamics of this distribution function is governed by the appropriate Boltzmann equation.
We are interested in the kinetic theory of chiral fermions obtained from quantizing relativistic Weyl fermion fields.
For simplicity, we restrict our discussion to the case of a single Weyl fermion field in this and the next section~\ref{sec:left}.
The generalization to the case of two Weyl fermions with opposite chiralities
(or equivalently, a single massless Dirac fermion) is straightforward and will be considered in section~\ref{sec:sphaleron}.

Quantizing a single right-handed Weyl fermion field, one obtains right-handed chiral particle and 
left-handed anti-particle.
With the assignment of a $U(1)$ charge $+1$ to the Weyl fermion field,
the particles (anti-particles) carry the charge $+1$ ($-1$).
We let these charges couple to a $U(1)$ gauge field (which we will call ``electromagnetism'') with a weak coupling strength $e$, which we assume to be non-dynamical\footnote{This assumption can be justified in a theoretical limit where $e\to 0$ while $eB$ kept fixed.}.
Let us denote the distribution functions of particles and anti-particles by
$f_{\pm}(t, \vx, \vp)$ where $\pm$ means
particle and anti-particle respectively.
The Boltzmann kinetic equation for $f_{\pm}(t,\vx,\vp)$ reads as:
\begin{equation}
\label{eq:p-L}
\frac{\pd f_{ \pm } }{\pd t} + \dot{\vx}\cdot\frac{\pd f_{\pm}}{\pd \vx}
+ \dot{\vp}_{\pm}\cdot\frac{\pd f_{\pm}}{\pd \vp} = \mathcal{C}_{\pm} [\, f_{+}, f_{-} \, ]\, ,
\end{equation}
where the equations of motion of the chiral particle and
anti-particle, expressing $(\dot{\vx},\dot{\vp}_{\pm})$ in terms of
($\bm x$, $\bm p$),
can be obtained from a semi-classical one-particle action derived from
path integral formulation of chiral fermions as shown in
Ref.~\cite{Stephanov:2012ki}. The crucial ingredient in the chiral kinetic theory is the
 Berry's phase and the corresponding geometric connection in momentum space arising from the projection of momentum dependent chiral spinor wave-function.
In the presence of a homogeneous magnetic field $\vB$ (without turning on an electric field), 
the resulting equations of motion can be written as
\begin{subequations}
  \label{eq:eom-L}
\begin{equation}
\label{eq:eomx-L}
\sqrt{G}\,\dot{\vx}
= \vv +(\vv\cdot\vb)\vB\, ,
\end{equation}
\begin{equation}
\label{eq:eomp-L}
\sqrt{G}\,\dot{\vp}_{\pm}
=
\pm\, \vv\times \vB\,  ,
\end{equation}
\end{subequations}
where
\be
\label{eq:b-def}
\vb \equiv 
 \frac{\hp}{2 |\vp|^2}\, ,
\qquad
\text{with}
\qquad
\hp =\frac{\vp}{|\vp|}\, ,
\ee
is the curvature of the Berry's connection in momentum space, 
\begin{equation}
\label{eq:G-def}
\sqrt{G} \equiv 1+ \vB\cdot\vb\, ,
\end{equation}
is the modified phase space density due to the interplay between the magnetic field and the Berry's curvature, and
\be
\label{eq:g-velocity}
\vv
=\frac{\pd \e}{\pd\vp}\,,
\ee
is the quasi-particle velocity with dispersion relation $\e=\e(\vp)$.
The equations of motion for particle and anti-particle are related by 
$\vb \to -\vb$ and $\vB\to -\vB$, so that in Eq.~\eqref{eq:eomx-L},
$\dot{\vx}$ and $\sqrt{G}$ 
are identical for particle and anti-particle.

In the presence of magnetic field, the Lorentz invariance dictates the existence of a spin-magnetic moment coupling in the dispersion relation \cite{Son:2012zy,Chen:2014cla},
\be
\label{eq:energy-B}
\e(\vp) = |\vp| -\frac{\vB\cdot\vp}{2|\vp|^2}\, ,
\ee 
which is essential for reproducing the correct value of chiral vortical effect at weak coupling~\cite{Chen:2014cla}.

The chiral kinetic theory is well justified only in the regime where the quasi-particle momentum $|\vp|$ is not much smaller than the temperature, so that
 the corrections to the dispersion relation from thermal effects can
 be neglected. In addition, we need to have $|\vp|\gg \sqrt{B}$ for a
 valid classical quasi-particle picture of motion in the magnetic
 field, instead of quantized Landau level picture.
This can also be seen in Eq.~\eqref{eq:G-def}, where 
 $\sqrt{G}$ as a modified phase space density will not make sense 
when $\sqrt{G}$ becomes negative in the region of momentum space $|\vp|\lesssim\sqrt{B}$.
In the region $|\vp|\lesssim\sqrt{B}$, quantum description is needed to describe the physics properly. Therefore our results based on chiral kinetic theory 
receive corrections from this quantum region. Nonetheless the size and
importance of the quantum region is suppressed for $B\ll T^2$  and the dominant contributions to the anomalous
transport phenomena (linear in $B$) come from the thermal momentum region $|\vp|\sim
T$. The corrections from the quantum regions are of higher order in
small $B$ expansion, i.e., we expect our results to be valid only up to linear order in $B$ while the quantum region contribution is of order $B^2$ or higher.
We emphasize that chiral kinetic theory allows us to study a frequency
and wave number region beyond the hydrodynamic regime, which is interesting even if we are restricted to the leading effects in small $B$ expansion.

\section{\label{sec:left} Chiral Magnetic Wave in hot Weyl gas}

In this section, we consider a high temperature plasma of a single species of right-handed Weyl fermion (that contains right-handed particles and left-handed anti-particles) in the presence of a homogeneous magnetic field $\vB$ to study the collective propagating mode of chiral charges originating from triangle anomaly:  the chiral magnetic wave. 
We assume that the temperature is high enough, $T\gg \sqrt{B}$ and
$T\gg\mu$, such that our results are well justified up to linear order
in $B/T^2$ as discussed in the previous section. This also justifies
our weak coupling picture in a QCD like gauge theory, although we will
not be specific about the interactions in our relaxation time
approximation. In section~\ref{sec:fermi}, we will discuss the
opposite limit $\mu\gg T$, of cold dense system of chiral fermions.

For simplicity, we shall consider the case of
neutral plasma $\mu=0$ where the charge fluctuation modes of
interest to us do not mix with the energy-momentum fluctuations.
We solve the kinetic equations \eq\eqref{eq:p-L} linearized in fluctuations of
 $f_{\pm} (t,\vx, \vp)$ from 
their equilibrium values,
\be
\label{eq:f0}
f^{0}(\e)
= \frac{1}{e^{\beta \e}+1}\,,
\ee
for both particle and anti-particle. 
Parameterizing the fluctuations with $h_{\pm}(t,\vx,\vp)$ as
\begin{equation}
  \label{eq:f-h}
  f_{\pm} (t,\vx, \vp)
=  f^{0}(\e)-\frac{\pd f^{0}(\e)}{\pd\e}\,h_{\pm} (t,\vx,\vp)\, ,
\end{equation}
and linearizing the collision term (noting that the
collision term vanishes for equilibrium distributions, ${\cal C}[f^{0},f^{0}]=0$),
\begin{equation}
  \label{eq:collision-linearized}
  {\cal C}_{\pm}[\,f_{+},f_{-}\,] =
-\frac{\pd f^{0}(\e)}{\pd\e}\,{\cal I}_{\pm}[\,h_{+},h_{-}\,] + {\cal O}(h^2)\,,
\end{equation}
with a functional linear operator ${\cal I}_{\pm}$ acting on $h_{\pm}$, the linearized kinetic equation
takes the following form in Fourier space,
\begin{equation}
  \label{eq:kinetic-linear-L}
  -i \o h_{\pm} +
\dot{\vx}\cdot\(i\vk \pm {\vB}\times \frac{\pd}{\pd\vp}\)h_{\pm}
 =
 {\cal I}_{\pm}[\, h_{+}, h_{-}\, ]\, ,
\end{equation}
with frequency-momentum $(\o,\vk)$.

In the next subsection \ref{sec:hydro-L}, we will derive the dispersion relation of chiral magnetic wave
in the hydrodynamic regime of low frequency and wave number limit in a
general way without using any explicit form of the collision operator
${\cal I}_{\pm}$ demonstrating the generality of the chiral magnetic
wave. In the subsection \ref{RTA}, we will consider the case of
relaxation time approximation for the collision term that allows us to
find some analytic solutions for finite frequency and wave number
outside of the hydrodynamic regime.

\subsection{Dispersion relation in hydrodynamic regime}
\label{sec:hydro-L}

Since the equilibrium distribution with an arbitrary chemical potential $\mu$,
\be
\label{eq:fermi-dirac}
f^{0}_{\pm}(\e)
= \frac{1}{e^{\beta(\e \mp \mu )} +1 }\, ,
\ee
should be a solution to the kinetic equation \eq\eqref{eq:p-L}, the linearized kinetic equation \eq\eqref{eq:kinetic-linear-L} must
have a static, homogeneous ($k=\o=0$) zero mode solution corresponding to an infinitesimal shift of chemical potential. With the definition of $h_{\pm}$ in \eq\eqref{eq:f-h},
this zero mode becomes $h_{+}=-h_{-}=\text{const}$ ($\vp$ independent), that is, ${\cal I}_\pm[1,-1]=0$ should be satisfied.
At a small but finite $\vk$,
the solution $h_{\pm}$ to \eq\eqref{eq:kinetic-linear-L} describes a
hydrodynamic mode, whose dispersion relation will be determined in the
following way. 

For future convenience, 
we define a linear functional, $\<\ldots\>$,
which can be thought of as averaging over the momentum space:
\be
\label{eq:average1}
\<\, \ldots \,\>\equiv
-\frac{2}{\chi}\int_{\vp}\sqrt{G} \frac{\pd f^{0}(\e)}{\pd\e}(\ldots)\,,
\ee
where \be
 \int_{\vp}\equiv \int \frac{d^3 \vp}{(2\pi)^3}\,,
 \ee
and
\begin{equation}
  \chi \equiv \(\frac{\pd n}{\pd \mu}\)_{\mu=0}
  = -2\int_{\vp} \sqrt{G} \frac{\pd f^{0}(\epsilon)}{\pd\epsilon}
= \frac{T^2}{6}+{\cal O}\(\frac{B^2}{T^2}\)
\, ,
\end{equation}
is the charge susceptibility at zero chemical potential. Note the normalization
$\<\, 1\,  \> = 1$.

Let us apply the operation of ``average'' defined in
\eq\eqref{eq:average1} to the equation \eq\eqref{eq:kinetic-linear-L}.
The charge conservation dictates that the collision term must satisfy
\begin{equation}
\label{eq:collision-integral-h-L}
\int_{\vp}\sqrt{G}\,
\(
\,{\cal C}_{+}[\,f_{+},f_{-}\, ]
-{\cal C}_{-}[\,f_{+},f_{-}\, ]\,
\)
=0\, ,
\end{equation}
for any distributions $f_\pm$, which implies
\be
\label{eq:h-condition}
\left<({\cal I}_+-{\cal I}_-)[h_{+},h_{-}]\right>=0\, ,
\ee
for an arbitrary fluctuation $h_{\pm}$.
It is easy to show that the Lorentz force term in \eq\eqref{eq:kinetic-linear-L} vanishes
upon averaging using the equation (\ref{eq:eomx-L}) for $\dot{\vx}$ and integrating by parts, so that
we have 
\be
-i\o \<h\> +i\vk\cdot\<\dot{\vx} h\>=0\, ,
\ee
after subtracting the two equations, where $h\equiv h_+-h_-$.
In general,
we need to know $h$ up to normalization to find the dispersion relation from the above equation.

However, since we know that $h=$ const when $\vk=0$ (the zero mode that is discussed above),
we expect $h=1+{\cal O}(\vk)$, and inserting this into the above gives
\be
\o = \vk \cdot \< \dot{\vx}\> +{\cal O}(\vk^2)\, . 
\ee
Using 
\eq\eqref{eq:eomx-L} for $\dot{\vx}$ to compute $\< \dot{\vx}\>$,
we see that the first term does not contribute as it is an integration of a total derivative, and the second term gives
\be
\label{xdoteq}
\<\dot{\vx}\> =-\frac{2}{\chi}\int_{\vp}\frac{\pd f^{0}(\e)}{\pd \e}
(\vv\cdot\vb)\vB=-{2\over\chi}\int_{\vp} \({\partial f^0\over\partial \vp}\cdot\vb\) \vB={2\over\chi}\int_{\vp} f^0\(\nabla_{\vp}\cdot\vb\) \vB
= \frac{\vB }{4\pi^2\chi}\,,
\ee
where we have used $\nabla_{\vp}\cdot\vb=2\pi\delta^{(3)}(\vp)$ and
$f^{0}(0)=1/2$.
This finally gives the dispersion relation $\o=\vv_{B}\cdot \vk+{\cal O}(\vk^2)$ with the group velocity 
\be
\vv_{B} = \frac{\pd\o}{\pd\vk} 
=\frac{\vB}{4\pi^2\chi}\, .
\ee
In contrast with the case of sound waves whose dispersion relation
is $\o=c_s|\vk|$ which is non-analytic in $\vk$,
the above chiral magnetic wave has a well-defined group velocity $\vv_B=(\pd\o(\vk)/\pd\vk)|_{\vk=0}$ at zero momentum, confirming the result in Ref.\cite{Kharzeev:2010gd}.
Remarkably,
the group velocity of chiral magnetic wave $\vv_{B}$
obtained from chiral kinetic theory
has no dependence on the details of $\e({\vp})$ or the collision operator ${\cal I}$.
Like sound waves,
it is given solely by a thermodynamic property (the charge susceptibility) of the system~\cite{Kharzeev:2010gd}.

\subsection{Relaxation time approximation}
\label{RTA}

In the previous subsection, 
we have determined the dispersion relation of chiral magnetic wave from chiral kinetic theory in the hydrodynamic regime.
However,
kinetic theory is a microscopic framework and contains more information than hydrodynamics.
That microscopic information is encoded in the details of the collision operator ${\cal I}_{\pm}$, which eventually determines
the dispersion relation beyond hydrodynamic regime. 
To see qualitative features of such dispersion relation beyond hydrodynamic regime with some analytic control, let us consider a simple example of collision operator in
the relaxation time approximation (RTA) (see also
Ref.\cite{Satow:2014lva} for another application of RTA in chiral kinetic theory):
\begin{equation}
\label{eq:BGK}
{\cal I}_{\pm}[\,h_{+}, h_{-}\,] =
- \frac{1}{\tau} \(\,h_{\pm} \mp \bar{h} \, \)
\end{equation}
with a relaxation time $\tau$, and
\begin{equation}
\label{eq:hbarL}
\bar{h}
\equiv \frac{1}{2}\<\, h_{+}-h_{-}\, \> \, ,
\end{equation}
from the charge conservation constraint (see \eqref{eq:h-condition}). 
$\bar{h}$ may be interpreted as a local fluctuation of chemical potential.

To simplify our analysis, let us consider only the case where $\vk$ is
parallel to $\vB$  and define $k$ via $\bm k = k\bm B/|\bm B|$.
In this case, 
$h_{\pm}$
 is independent of the azimuthal angle in the transverse plane perpendicular to $\vB$, and the Lorentz force term in Eq.~\eqref{eq:kinetic-linear-L} vanishes.
Then, Eq.~\eqref{eq:kinetic-linear-L} becomes
\be
\label{eq:h-RTA1}
\(-i\,\o + i\,\dot{\vx}\cdot\vk +\tau^{-1}\,\)h_{\pm}
=
\pm \frac{\bar{h}}{\tau} \, .
\ee
For real $k$ as long as ${\rm Im} (\omega)\neq -\tau^{-1}$, we can solve \eqref{eq:h-RTA1} to obtain
\be
\label{eq:h0haL}
h_{\pm}
= 
\pm \frac{\tau^{-1}}{ -i\o\, +i\, \dot{\vx}\cdot\vk + \tau^{-1} }\, \bar{h}\, .
\ee
The mode $\omega = \bm{\dot x}\cdot\bm k - i \tau^{-1}$
corresponds to quasiparticle excitations. We, however, are interested
in the collective modes which are given by \eq\eqref{eq:h0haL}.
Substituting \eq\eqref{eq:h0haL} into \eq\eqref{eq:hbarL},
we get a self-consistent integral equation that can be used to determine the dispersion relation,
\be
\label{eq:AL}
A\equiv \left<\,\frac{\tau^{-1}}{-i\o\, +i\, \dot{\vx}\cdot\vk + \tau^{-1}\,}\right>
=1\,.
\ee
Recall the definition of $\< \ldots \>$ in \eqref{eq:average1}.

We will solve the above integral equation up to linear order in $\vB$ that our kinetic theory is justified.
Expanding $\o(k)$ in $\vB$ we can write
\footnote{Since $\omega$ is parity even while $\bm B$ is an axial
  vector,
the term  $\omega_\anom(k)$ breaks parity, and is therefore labeled as ``anomalous''.}
\be
\label{eq:ob-ex}
\o(k) = \o_\norm(k) 
+ \o_{\anom}(k)+\cdots\, ,
\ee
where $\omega_\norm(k)={\cal O}(\bm B^0)$ and $\omega_\anom(k)={\cal
  O}(\bm B^1)$. At $\bm B=0$ Eq.~(\ref{eq:AL}) for
$\omega_\norm(k)$ 
is easily solved giving~\footnote{
It should be noted that the solution to $A=1$ exists only for $k\tau<\pi/2$
because the function $A(\omega,k)$ in Eq.~(\ref{eq:AL}) maps the whole
complex plane of $\omega$ into a vertical strip $|{\rm
  Re}A|<\pi/(2k\tau)$. For $k\tau>\pi/2$, the response of the gas is
determined by
quasiparticle excitations.
}

\begin{equation}
\label{eq:omega-norm}
\o_{\norm}(k) = 
-i\(\, \tau^{-1} - \,k \cot(k\,\tau)\,\)\, .
\end{equation}
Indeed, $\o_{\norm}(k)$ is the known dispersion relation of charge diffusion under the relaxation time approximation. 
For $k \tau \ll 1$, $\o_{\norm}(k) $ can be expanded as
\begin{equation}
\label{eq:norm-hydro}
\o_{\norm}(k) \approx -i\(\,\frac{1}{3}(k\,\tau)^2 +\frac{1}{45}(k\,\tau)^4+\cdots\,\)\tau^{-1}\, ,
\end{equation}
and the first term gives the normal diffusion constant under the relaxation time approximation,
\be
\label{eq:D-norm}
D_{\norm} = \frac{1}{3}\tau \, .
\ee

The equation for $\omega_\anom(k)$ can be also solved analytically giving 
\begin{equation}
\label{eq:o-anom-L}
\o_{\anom}(k)
= {B\over 4\pi^2 \chi} k=v_B k \, ,
\end{equation}
confirming existence of the chiral magnetic wave in the hydrodynamic regime.
In the RTA this result for
the anomalous part of the dispersion relation turns out (accidentally)
to be exact beyond the hydrodynamic
regime.
More importantly,
$\o_{\anom}(k)$ is purely real and hence non-dissipative. This should
remain true also outside the RTA, i.e., for a more realistic collision operator.

\section{Plasma of fermions with two opposite chiralities}
\label{sec:sphaleron}

Generalizing the previous section, let us consider the more realistic
case of two Weyl fermions of opposite chiralities (equivalently, one
massless Dirac fermion), which is an approximation to the QCD with
light quarks.  In QCD, topological sphaleron transitions can change
the axial charge defined as a difference between the charges of the
left-handed Weyl fermion and the right-handed one.  This
non-perturbative sphaleron dynamics leads relaxation of the axial
charge imbalance, whose time scale is in general different from the
time scale of the relaxation that achieves local thermal equilibrium
at fixed chirality.  With the weak coupling QCD results in mind,
we will assume that the inter-chiral relaxation time (which we denote
as $\tau_s$) provided by sphaleron transitions is much longer than the
thermal relaxation time $\tau$: $\tau_s\gg \tau$.

Based on the separation of these two scales,
we can expect the following picture:
On the time scale longer than the relaxation time $\tau_s$,
sphaleron transitions are effective and the axial charge conservation is heavily violated.
What remains conserved is the total vector charge of the system
defined as a sum of left-handed and right-handed chiral charges.
We expect to have a simple diffusion mode of this vector charge as usual.
However,
at time scales shorter than the sphaleron relaxation time $\tau_s$,
but still
longer than the thermal relaxation time $\tau$,
the axial symmetry, and hence the full chiral symmetry $U(1)_L\times U(1)_R$,
remains approximately conserved,
and we expect to have the triangle anomaly in action to give us the features of chiral magnetic waves.
There would be two propagating modes along the direction of magnetic field moving in opposite directions,
corresponding to chiral magnetic waves of left-handed and right-handed Weyl fermions.
In summary,
as we vary the time  scale (i.e., frequency or wave number) of the probe, we expect to see a transition from a simple diffusive behavior to the chiral magnetic wave.
We will verify such a picture in this section explicitly 
in the relaxation time approximation.

The chiral kinetic equation in the case of two Weyl fermions with opposite chiralities is a simple generalization of our previous kinetic equation Eq.~\eqref{eq:p-L}:
\be
\label{eq:kinetic-LR}
\( \, \frac{\pd}{\pd t} + \dot{\vx}_{\chi}\cdot\frac{\pd}{\pd\vx}
+ \,\dot{\vp}_{\chi,\pm}\cdot\frac{\pd}{\pd\vp}
\, \) f_{\chi,\pm} = \,{\cal C}_{\chi, \,\pm} [\, f_{\chi',+},f_{\chi',-} \,  ]\,,
\ee
where $\chi, \chi'=\text{R,L}$ label the chirality and $\pm$ denotes the vector $U(1)$ charge (particle or anti-particle) as before.
The equations of motion \eq\eqref{eq:eom-L}
in the presence of magnetic field
can also be generalized accordingly to
\begin{subequations}
  \label{eq:eom}
\begin{equation}
\label{eq:eomx}
\sqrt{G_{\chi}}\,\dot{\vx}_{\chi}
= \vv_{\chi} + \, (\vv_{\chi}\cdot\vb_{\chi})\, \vB\, ,
\end{equation}
\begin{equation}
\label{eq:eomp}
\sqrt{G_{\chi}}\,\dot{\vp}_{\chi,\pm}
= \pm \, \vv_\chi\times \vB\,   ,
\end{equation}
\end{subequations}
where we define $b_{R}=b, b_{L}=-b$, 
\begin{equation}
\sqrt{G_{\chi}} = 1+ 
\, \vB\cdot\vb_{\chi}\, ,
\end{equation}
and $\vv_\chi=\pd \epsilon_{\chi}/\pd \vp$ with
$\epsilon_{\chi}=|\vp|- (\vB\cdot\vb_\chi)|\vp|$.
The above generalization can easily be deduced by simple flipping of the sign of the Berry curvature for opposite chirality of fermions.
As before,
we linearize the above kinetic equation by expanding around
the equilibrium distribution $f^{0}_\chi\equiv 1/ (e^{\b \e_\chi}+1)$,
\begin{equation}
  \label{eq:f-h-LR}
  f_{\chi, \pm} (\vp, \vx, t) = f^{0}_\chi
-\frac{\pd f_\chi^{0}}{\pd\epsilon_\chi}\,\,h_{\chi, \pm} (\vp,\vx, t)\, ,
\end{equation}
and introducing the linearized collision operators ${\cal I}_{\chi,\pm}$,
\begin{equation}
  {\cal C}_{\chi,\pm}[\,f_{\chi', +},f_{\chi',-}\,] =
 -\frac{\pd f_\chi^{0}}{\pd\epsilon_\chi}\,\,{\cal I}_{\chi,\pm}[\,h_{\chi',+}, h_{\chi',-}\,] + {\cal O}(h^2)\, .
\end{equation}

We consider the relaxation time approximation for the collision operators ${\cal I}_{\chi,\pm}$, assuming that the two relaxation dynamics
corresponding to thermal equilibrium within the same chirality
and inter-chiral sphaleron transitions 
are independent of each other. 
The former dynamics will introduce a linearized collision operator that is similar to the one discussed in the previous section,
\be
{\cal I}_{\chi,\pm}^{\rm thermal}
[\, h_{\chi,+},h_{\chi,-}\,]=- \frac{h_{\chi,\pm} \mp\bar{h}_\chi}{\hat\tau}\,,
\ee
with
\be
\label{eq:hLR}
\bar{h}_{\chi} \equiv 
\frac{1}{2}\<\,h_{\chi,+}- h_{\chi,-}\,\>_{\chi} 
\, ,
\ee
where the average operation $\left<\ldots\right>_\chi$ is defined for each chirality as before (see \eqref{eq:average1}).
The $\bar{h}_{\chi}$ correspond to local fluctuations of chiral
chemical potentials.
On the other hand, the inter-chiral transition dynamics will induce a collision operator
\be
\label{eq:sphaleron}
{\cal I}_{\chi,\pm}^{\rm inter-chiral}
[\, h_{\chi',+},h_{\chi',-}\,]=-\frac{1}{\tau_{s}}\(h_{\chi,\pm}\mp\bar h\)\,,
\ee
with
\be
\bar{h} 
\equiv\frac{1}{2}\(\, \bar{h}_{R}+\bar{h}_{L}\,\)
\, ,
\ee
which corresponds to a local fluctuation of vector chemical potential.

Restricting ourselves to the case of $\vk$ parallel to $\vB$ as before
($\vk=k\bm B/|\bm B|$), the linearized kinetic equation with the above collision operators 
read as
\bes
\label{eq:hh50}
\be
\[- i\o +i \dot{\vx}_{R}\cdot\vk +\tau^{-1}
\,\]\, h_{R,\pm}
=\pm \[\, \tau^{-1} \bar{h}_{R} \,- \tau^{-1}_s\, \bar{h}_5
\,\] \, ,
\ee
\be
\[- i\o +i \dot{\vx}_{L}\cdot\vk +\tau^{-1}
\,\]\, h_{L,\pm}
=\pm \[\, \tau^{-1} \bar{h}_{L} \, + \tau^{-1}_s\, \bar{h}_5
\,\]\, ,
\ee
\ees
where we defined $\bar{h}^5$  as
\be
\bar{h}^{5} 
\equiv \frac{1}{2}\(\,\bar{h}_R - \bar{h}_L\,\)\,,
\ee
which corresponds to a fluctuation of the axial chemical potential, 
and introduced $\tau\equiv(\hat\tau^{-1}+\tau^{-1}_s)^{-1}$.
Since we consider $\tau_s\gg\tau$, we have
$\tau\approx\hat\tau$. Solving for $h_{\chi,\pm}$ we find
\bes
\be
\label{eq:hLR0}
 h_{R,\pm}
 =\pm
 \frac{
\tau^{-1}
}
{
\[\,- i\o +i \dot{\vx}_{R}\cdot\vk +\tau^{-1}
\,\]
}
\left(\,\bar{h}_{R}\,-\left({\tau\over\tau_{s}}\right)\bar{h}_5\,\right)
 \, ,
\ee
\be
\label{eq:hLR1}
 h_{L,\pm}
 =\pm
 \frac{
\tau^{-1}
}
{
\[\,- i\o +i \dot{\vx}_{L}\cdot\vk +\tau^{-1}
\,\]
}
\left(\,\bar{h}_{L}\,+\left({\tau\over\tau_{s}}\right)\bar{h}_5\,\right)
 \, ,
\ee
\ees
Substituting the above into \eq\eqref{eq:hLR},
we obtain a coupled matrix equation
\be
\label{eq:h0ha-matrix}
\(
\begin{array}{cc}
  (1 -A) \,& -A_5\left(1 -\left({\tau/ \tau_{s}}\right)\right)\\
   -A_5 \,& 1-A\left(1-\left({\tau/\tau_{s}}\right)\right)
\end{array}
\)
\(
\begin{array}{c}
  \bar{h}\\
  \bar{h}^{5}
\end{array}
\)
= 0 \, ,
\ee
where $A \equiv \frac{1}{2}(A_R+A_L)$ and $A_5 \equiv \frac{1}{2}(A_R-A_L)$ with
\be
A_{\chi}\equiv \left\langle\,\frac{\tau^{-1}}{- i\o + i\dot{\vx}_\chi\cdot \vk+\tau^{-1}}\,\right\rangle_{\chi}\, .
\ee
The dispersion relation is then determined by zeros of the determinant of the matrix in \eq\eqref{eq:h0ha-matrix}, which is a quadratic equation involving $A$ and $A_5$.
From this, we will obtain two branches of dispersion relations,
$\o_{\pm}(k)$.

Before presenting our numerical results,
it is useful to examine various limits where analytic solution is possible.
We first note that in the infinite $\tau_s$ limit, i.e. with no inter-chiral transitions,
the equation for vanishing determinant becomes
\begin{equation}
\label{eq:A-A5-tau-inf}
(1-A_R)(1-A_L)=0\, ,
\end{equation}
which gives two independent branches of solutions: $A_R=1$ and $A_L=1$.
Recalling that the dispersion relation for the case of a single
right-handed Weyl fermion was obtained from $A_R=1$ in the previous section
(see \eqref{eq:AL}), 
we see that this limit correctly reproduces two separate chiral magnetic waves from each chirality.

Let us now consider a finite $\tau_s$ with $\o,v_Bk\ll \tau^{-1}$ (recall that we already consider $\tau_s^{-1}\ll\tau^{-1}$).
From the equation (see \eqref{xdoteq}),
\be
\langle\,\dot{\vx}_{R}\cdot\vk\,\rangle_{R}= v_B k\,,\qquad 
\langle\,\dot{\vx}_{L}\cdot\vk\,\rangle_{L}= - v_B k\,,
\ee
we have the expansion \be
A\approx 1+ i\o\tau +{\cal O}\left((\omega\tau)^2,(k\tau)^2\right) \, ,\qquad
A_5 
\approx -iv_Bk \tau + {\cal O}\left((\omega\tau)^2,(k\tau)^2\right) \, ,
\ee
from which the equation for the dispersion relation becomes
\be
\label{eq:omega-pm}
\o_{\pm}(k) =-i\,
\frac{1 \pm \sqrt{1-(2 v_{B} k\tau_s)^2}}
{2\tau_s} \, .
\ee

There are two qualitatively different regimes depending on whether
$kv_B\tau_s\ll 1$ or $kv_B\tau_s\gg 1$. The former is the regime where
the inter-chiral transitions play an important role so that axial charge conservation is heavily violated, while the latter is the case where one can neglect inter-chiral transitions and the two chiral magnetic waves with opposite chirality can be found. 
To see these explicitly, in the first case,
$kv_B\tau_s\ll 1$,
expanding  \eq\eqref{eq:omega-pm} gives one solution
$\o_{+}(k)=-i\tau^{-1}_s+ \mathcal{O}(k^2)$,
which is not a hydrodynamics mode: this mode simply corresponds to a relaxation of axial charge due to inter-chiral transitions. The other solution is
\be
\label{eq:omega-p}
\o_{-}(k) = -i v^2_B\tau_{s}  k^2 + \cdots
\equiv -i D_{\anom} k^2 + \cdots\,,
\ee
that features a diffusion behavior with a diffusion constant 
\be
\label{eq:D-anom}
D_{\anom} = \cmw^2\tau_{s}\, ,
\ee
which corresponds to the dynamics of the conserved vector charge of the system.

We note that the diffusion constant for the vector charge in the presence of magnetic field is not determined by the thermal relaxation time $\tau$, but by the dynamics of chiral anomaly. This can be understood as follows.
Via the Einstein relation,
one may relate $D_{\anom}$ to the electric conductivity
\be
\label{eq:s-anom}
\s_{\anom} = \chi_V D_{\anom}
=\chi_V \cmw^2\tau_s\, ,
\ee
where $\chi_V$ is the vector charge susceptibility which is $\chi_V=\chi_R+\chi_L=T^2/3$.
To understand this, consider a homogeneous electric field $\vE$ applied parallel to the magnetic field. Due to chiral anomaly, there is a constant generation of axial charge $n_5$ with the rate $1/(2\pi^2)\vE\cdot\vB$, and the time evolution equation of $n_5$,
\be
\dot{n}_5=-{1\over \tau_s}n_5+{1\over2\pi^2} EB\,,
\ee
gives the stationary value $n_5=\tau_sEB/(2\pi^2)$ or, equivalently, the axial chemical potential $\mu_5=\tau_sEB/(2\pi^2 \chi_V)$. With the magnetic field, this induces the chiral magnetic current
\be
J={\mu_5\over 2\pi^2}B={\tau_s B^2\over (2\pi^2)^2\chi_V} E=\chi_V \cmw^2\tau_s E\equiv\sigma_{\anom} E\, ,
\ee
which reproduces the above conductivity (we used $v_B=B/(4\pi^2\chi)=B/(2\pi^2\chi_V)$). 
Indeed,  \eq\eqref{eq:s-anom} is in agreement with the result of Ref.\cite{Son:2012bg}
which obtained the anomalous conductivity $\s_{\anom}$ in a zero temperature,
finite density system using the linear response theory, which also has been checked in holography~\cite{Landsteiner:2014vua}.

Alternatively, $D_{\anom}$ may also be interpreted as a diffusion
constant arising from the collective motion (drift) of chiral charges
by chiral magnetic wave of the speed $v_B$ with a relaxation rate
$\tau_s^{-1}$.

On the other hand, in the opposite regime $kv_B\tau_s\gg1 $, it is
easy to see that inter-chiral transitions do not play a role and
\eq\eqref{eq:omega-pm} produces two chiral magnetic waves moving in
opposite directions. We see these features in our numerical solutions presented below.

\begin{figure}[t]
  \centering
  \subfigure[]{
    \label{fig:diff-tau-dis-re}
	\includegraphics[width=.45\textwidth]{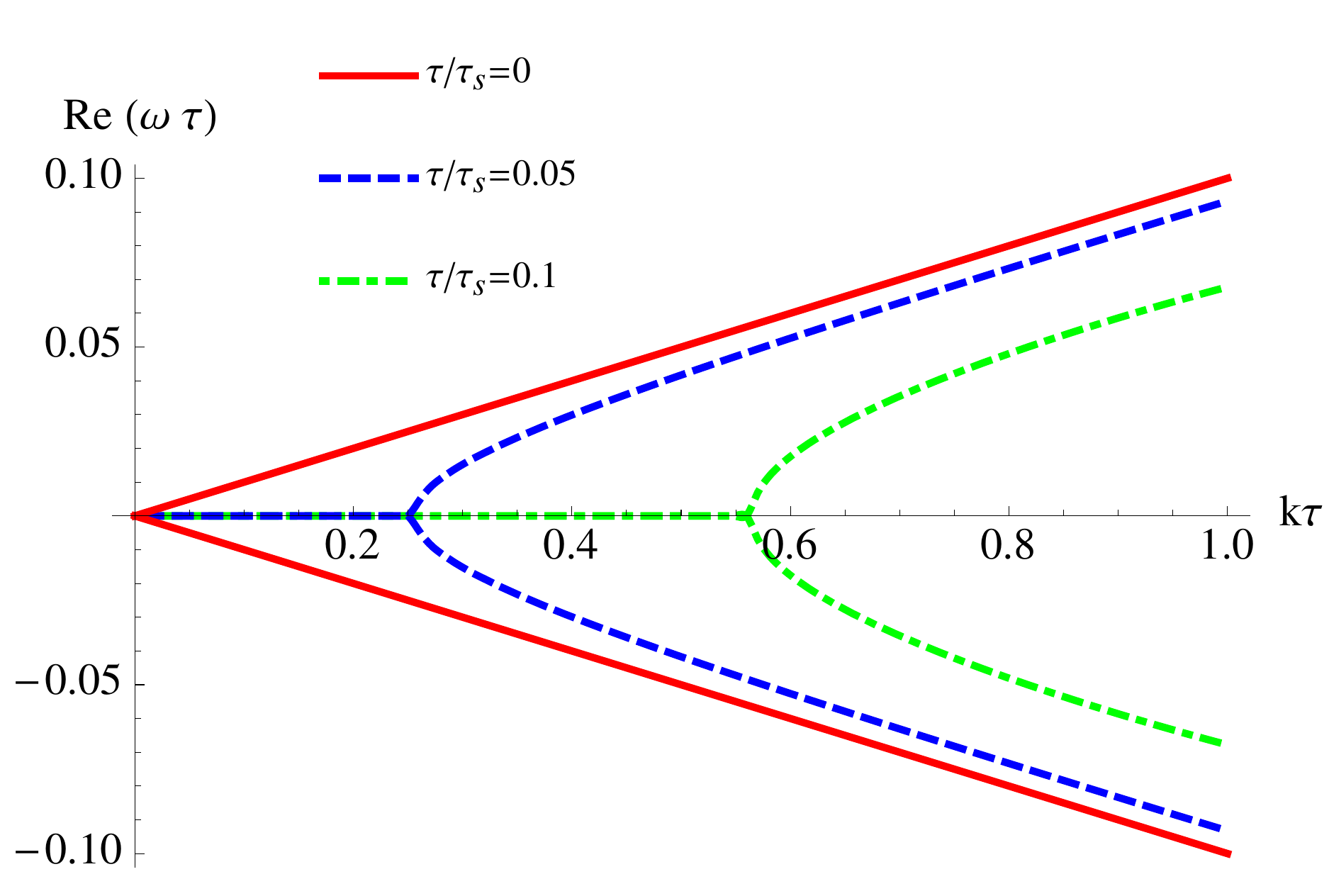}
       }
	\subfigure[]{
          \label{fig:diff-tau-dis-im}
	\includegraphics[width=.45\textwidth]{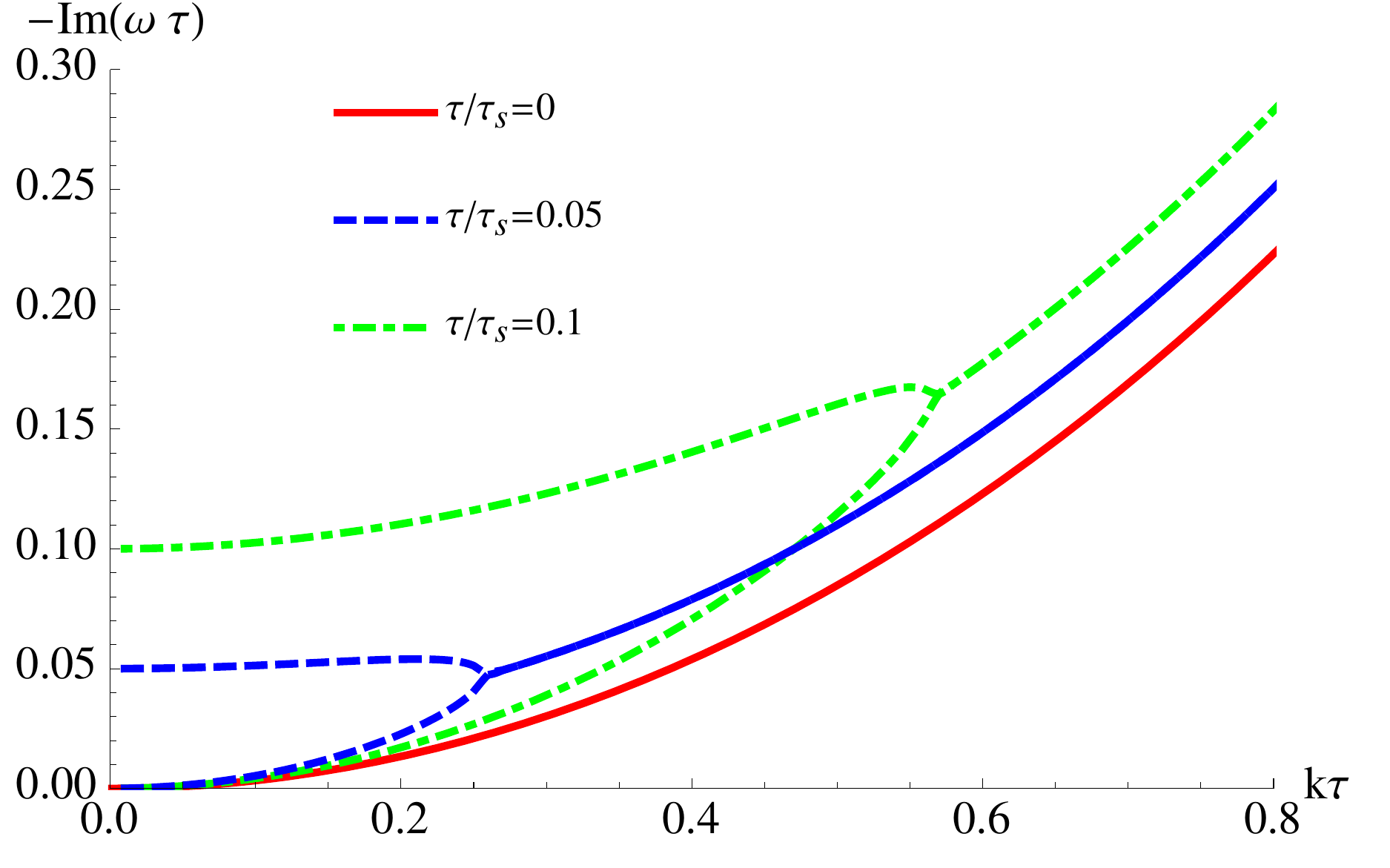}
           }
	\caption{
          \label{fig:diff-tau}
          Left: Re\,$\o_{-}(k)$ obtained from \eq\eqref{eq:h0ha-matrix} at $\tau/\tau_s = 0$ (red), $\tau/\tau_s = 0.05$ (blue) and
          $\tau/\tau_s=0.1$ (green).
 Right: $-{\rm Im\,} \o_{-}(k)$ obtained from \eq\eqref{eq:h0ha-matrix} with the same values of $\tau/\tau_s$.          In both plots, the speed of chiral magnetic wave is fixed at $\cmw = 0.1$.
	}
\end{figure}
In FIG.~\ref{fig:diff-tau},
we plot the dispersion relations $\o_{\pm}(k)$ determined from \eq\eqref{eq:h0ha-matrix}
with $\cmw = 0.1$, $\tau/\tau_s=0$(red), $\tau/\tau_s = 0.05$(blue) and $\tau/\tau_s=0.1$(green), which feature the above discussed transition from the diffusive mode in $kv_B\tau_s\ll1 $ to the chiral magnetic waves in $kv_B\tau_s\gg1 $.
As we have anticipated earlier, for small $k$, both $\omega_\pm$ are purely imaginary, and
$\o_{+}(k)$
is a non-hydrodynamic mode with a finite imaginary part even at $k=0$, and $\o_{-}(k)$ is a diffusive mode
with the diffusion constant given by $D_{\anom}+D_{\norm}$ where $D_{\norm}=\tau/3$ (\eqref{eq:D-norm}) (see the right plot).
Once $k$ reaches a critical value, $k\approx (2v_B\tau_s)^{-1}$
according to \eq\eqref{eq:omega-pm}, the two modes $\omega_\pm(k)$ 
meet with each other, and then split again into two modes now having non-zero 
real part (see the left figure), and each mode starts to have a non-zero group velocity featuring the chiral magnetic waves moving in opposite directions.
This group velocity becomes close to the asymptotic chiral magnetic wave velocity $v_B$ as $k$ increases.
The (negative) imaginary part of $\o_{\pm}(k)$ are well described by $D_{\norm} k^2$
as $k$ increases.

We note that these features are also found in a recent work in AdS/CFT correspondence~\cite{Jimenez-Alba:2014iia} where the authors introduced axial charge breaking (corresponding to inter-chiral transitions we consider) via a mass term for the axial gauge field in the holographic 5 dimensional space-time.

\section{Collective waves in cold chiral Fermi liquid}

\label{sec:fermi}
In this section, we consider a different temperature/density regime:
a cold dense system of chiral Fermions with $\mu\gg T$ and $\mu T\gg B$.
The second condition is needed to ensure that the Landau quantization
can be neglected within the thin shell of width $T$ 
around the Fermi surface where the quasi-particles can scatter.
The chiral kinetic theory based on weakly interacting quasi-particles
around the Fermi surface has been discussed recently
\cite{Son:2012wh,Son:2012zy}, 
and we would like to study some of the collective modes of the system in the presence of the magnetic field. 
In particular,
we show the existence of the chiral magnetic wave in this system.
Moreover,
in the interacting cold Fermi liquid,
there is another type of collective propagating mode -- the Landau's
zero sound. 
We will show how the dispersion relation of the zero sound mode is modified by the chiral anomaly
and study the transition from the chiral magnetic wave  in the
hydrodynamic regime to the 
zero sound mode in the collisionless regime as one increases the wave number $k$.

\newcommand\np{{n}}
\newcommand\dedp{{\bm v}}
\newcommand\mdedx{\bm{{\cal E}}}

At low temperature, 
we only need to consider quasi-particle excitations near the Fermi surface,
and for simplicity we will restrict our discussions to the case of a single right-handed Weyl fermion. 
The chiral kinetic equation for these quasi-particles takes a similar
form to the previous one in \eq\eqref{eq:p-L} 
(c.f.~ Ref.~\cite{Son:2012wh}), 
\begin{equation}
\label{eq:kinetic-particle-0}
\frac{\pd \np }{\pd t} + \dot{\vx}\bm\cdot\frac{\pd \np}{\pd \vx}
+ \dot{\vp}\bm\cdot\frac{\pd \np}{\pd \vp} = 
\mathcal{C}[ \np]\, ,
\end{equation}
where the distribution function is now denoted by $\np$ according to the prevailing convention,
and the equations of motion for $(\vx,\vp)$,
in the absence of external electric field,
are given by \cite{Son:2012wh}:
\bes
\label{eq:eom-zero}
\begin{equation}
\label{eq:eom-x}
\sqrt{G}\,\dot{\vx} =
\dedp
+\vB\, \left(\dedp\cdot\vb\right)
+\mdedx\times \vb \, ,
\end{equation}
\begin{equation}\label{eq:eom-p}
\sqrt{G}\,\dot{\vp} =
\mdedx
+\dedp\times \vB 
+\( \mdedx\cdot\vB \) \vb
\, ,
\end{equation}
\ees
where 
$\sqrt{G}=1+\vB\cdot\vb$, $\dedp=\pd \e/\pd\vp$, $\mdedx=-\pd\e/\pd\vx$
and $\e=\e(\vp)$ is the energy of a quasi-particle at momentum $\vp$.
Importantly, 
the quasi-particle energy $\e(\vp)$ is also a functional of $n$ due
to interparticle interactions. In particular, the interaction is
responsible for the effective force $\mdedx$ driven by
inhomogeneity of the particle density $n$: $\mdedx=-(\delta\e/\delta
n)\circ(\pd n/\pd\vx)$.

\newcommand\eeq{\e_0}
\newcommand\npeq{\np_0}

We now linearize Eq.~(\ref{eq:kinetic-particle-0}) in
fluctuations of $n$ around the equilibrium Fermi-Dirac distribution
\begin{equation}
  \label{eq:n-eq}
   \np_{0} = \frac{1}{e^{\beta(\e_{0}-\mu)} + 1}\, ,
\end{equation}
with the equilibrium dispersion relation $\epsilon_{0}=\e_0(\vp)$.
We expand $n(t,\vx,\vp)$, and functionals ${\cal C}[\np]$ and
$\e=\e(\vp)[\np]$, defining function $h(t,\vx,\vp)$ 
\begin{equation}\label{eq:h-def}
  \np = \npeq + \frac{\pd  \npeq}{\pd \mu} h;
\end{equation}
linearized collision
functional ${\cal I}[h]$ (noting that collision
integral vanishes in equilibrium ${\cal C}[\npeq]=0$)
\begin{equation}\label{eq:I-def}
{\cal C} [\np] = 0 + \frac{\pd \npeq}{\pd \mu}\ {\cal I}[h] + \OO(h^2);
\end{equation}
and linearized Landau interaction functional ${\cal F}[h]$ 
\begin{equation}\label{eq:F-def}
  \e(\vp)[\np] = \eeq(\vp) + {\cal F}[h] + \OO(h^2).
\end{equation}
The conventional choice of the definition of the function $h$ in
Eq.~(\ref{eq:h-def}) is such that $h=\textrm{const}$ describes a
change of the distribution due to a shift of the chemical potential
$\mu$ by $h$. We also note that in the zero-temperature limit
($\beta\to\infty$) the coefficient
$\pd\npeq/\pd\mu\to\delta(\eeq-\mu)$, i.e., the fluctuations of $n$
are constrained to a thin shell around the Fermi surface.

As discussed in Ref.~\cite{Son:2012zy} from the point of view of quantum field theory,
\eq\eqref{eq:kinetic-particle-0}
and \eq\eqref{eq:eom-zero} may be derived from high density effective theory in which
the chemical potential $\mu$ is much larger than the excitation energy
of quasi-particles of order $T$ and $\sqrt{B}$. 
Since the derivation is only up to a linear order in $B$, 
we will keep our results only up to this order for consistency.

Expanding the second term on the l.h.s. of the kinetic equation
(\ref{eq:kinetic-particle-0}) using equation of motion
(\ref{eq:eom-x})
we find
\begin{equation}\label{eq:xdot-h}
 \bm {\dot x\cdot} \frac{\pd \np}{\pd \vx}
= \bm{\dot
  x}_0\cdot
\frac{\pd h}{\pd \vx}\frac{\pd  \npeq}{\pd \mu} 
+\OO(h^2);
\end{equation}
where we denote by $\bm{\dot
  x}_0$ full quasiparticle velocity in equilibrium
\begin{equation}
  \label{eq:xdot-0}
  \bm{\dot x}_0 \equiv 
\frac1{\sqrt G}\,(\bm v_0 + \bm B (\bm {v}_0\bm{\cdot b})) \,,
\end{equation}
with $\bm v_0 = \pd\e_0/\pd \vp$ being the normal particle velocity in
equilibrium. We linearize the third term on the l.h.s. of
Eq.~(\ref{eq:kinetic-particle-0}) using equation of
motion~(\ref{eq:eom-p})
and definitions (\ref{eq:h-def}), (\ref{eq:F-def}) and (\ref{eq:xdot-0})
\begin{equation}
  \label{eq:pdot-h}
    \bm {\dot p\cdot} \frac{\pd \np}{\pd \vp}
=\left[-\bm{\mdedx\cdot\dot x}_0
+  (\bm{\dot x}_0\bm{\times B})\bm\cdot\frac{\pd}{\pd \vp}(1+{\cal F})[h]
\right]\frac{\pd  \npeq}{\pd \mu}
+\OO(h^2)\,.
\end{equation}
We used the fact that $\mdedx=\OO(h)$ since, by definition,
$\mdedx=-\pd{\cal F}[h]/\pd\vx + \OO(h^2)$ and applied the relation
$\pd \npeq/\pd \vp = -\dedp_0 (\pd \npeq/\pd\mu)$.

Substituting Eqs.~(\ref{eq:xdot-h}) and~(\ref{eq:pdot-h}) into
Eq.~(\ref{eq:kinetic-particle-0}) one arrives at the linearized
kinetic equation, which after Fourier transform from $\vx$ to $\vk$ 
takes  the form
\begin{equation}\label{eq:kinetic-Landau-linearized}
  -i\omega h +\bm {\dot x}_0\bm\cdot\left( 
i\bm k + \bm B\bm\times\frac{\pd
    }{\pd\vp}\right) (1+{\cal F})[h] = {\cal I}[h]\,.
\end{equation}

\subsection{CMW at small wave number}

Since the Fermi-Dirac distribution $\npeq$ in Eq.~(\ref{eq:n-eq}) is a
solution to the kinetic equation (\ref{eq:kinetic-particle-0})
for any $\mu$, the linearized kinetic
equation~(\ref{eq:kinetic-Landau-linearized}) must have a zero mode at
$\bm k = \o=0$. In the noninteracting Fermi gas, that mode would be
simply $h=\textrm{const}$ ($\vp$-independent), 
since $h$, as defined in Eq.~(\ref{eq:h-def})
 coincides with the shift of the chemical potential. Interactions,
 however, cause the change of the energy $\e_0(\vp)$ to be
 different as well, and thus the two distributions at chemical
 potentials $\mu$ and $\mu'$ have different functions $\e_0(\vp)$ and 
$\e_0'(\vp)$ in the Fermi-Dirac distribution Eq.~(\ref{eq:n-eq}). The
difference between two such infinitesimally close 
distributions defines $\delta h$ -- the zero
mode, according to Eq.~(\ref{eq:h-def}),
\begin{equation}
  \label{eq:h0}
  \frac{\pd \npeq}{\pd\mu}\delta h \equiv \npeq'-\npeq
= \frac{\pd \npeq}{\pd\mu} (\delta\mu - \delta\eeq)\,.
\end{equation}
Since $\delta\eeq\equiv\eeq'-\eeq={\cal F}[\delta h]$, we find that even if
$\delta h$ could depend on $\vp$, the functional $\delta h+{\cal
  F}[\delta h]$ must
be $\vp$-independent, or
\begin{equation}
  \label{eq:h+F=mu}
  (1+{\cal F})[h_0]=1,
\end{equation}
where we defined (normalized) the zero mode as the ratio of two
infinitesimal quantities: $h_0\equiv \delta h/\delta\mu$. In the
absence of interaction, ${\cal F}=0$, the zero mode is $h_0=1$.

From the fact that ${\cal C}[\npeq']={\cal C}[\npeq]=0$ and the
definition of ${\cal I}$ in Eq.~(\ref{eq:I-def}) it follows that
\be
{\cal I}[h_0] = 0\, .
\ee 
The left hand side of the kinetic equation
(\ref{eq:kinetic-Landau-linearized}) also vanishes for $h=h_0$ due to
Eq.~(\ref{eq:h+F=mu}) at $\o=\vk=0$.

At small but non-zero $\bm k$, the solution $h$ to
Eq.~(\ref{eq:kinetic-Landau-linearized}) describes a hydrodynamic
mode, whose dispersion relation we can determine in the following way.

For further convenience we define a linear
functional, $\langle\ldots\rangle$, which can be thought of as
averaging over the momentum space
\begin{equation}
  \label{eq:average}
  \langle h \rangle \equiv 
 \frac1{\chi_0}
{\int_{\vp}\sqrt G\, \frac{\pd\npeq}{\pd\mu}\, h}\,,
\qquad\mbox{where}\qquad
\chi_0 \equiv
\int_{\vp}\sqrt G\, \frac{\pd\npeq}{\pd\mu}
\end{equation}
is the susceptibility of a free Fermi gas (neglecting integraction).

Let us apply operation of ``averaging'' defined in Eq.(\ref{eq:average})
to the equation~(\ref{eq:kinetic-Landau-linearized}).
Conservation of charge means $\int_{\vp} \sqrt G {\cal C}[n]=0$ for any
$n$  and therefore requires that
$\langle {\cal I}[h]\rangle=0$ for all $h$. The last term on the
r.h.s. of Eq.~(\ref{eq:kinetic-Landau-linearized}) also vanishes upon
such averaging, because 
\begin{equation}
  \label{eq:1}
 \int_{\vp} \sqrt G \frac{\pd\npeq}{\pd\mu}  \bm {\dot x}_0\bm\cdot\left( \bm B\bm\times\frac{\pd
    }{\pd\vp}\right) (1+{\cal F})[h]
=
\int_{\vp} \frac{\pd }{\pd\vp} \bm\cdot
\left(\frac{\pd\npeq}{\pd\vp}\bm{\times B} (1+{\cal F})[h]\right)=0
\end{equation}
is an integral of a total derivative.
Thus we obtain, for any $h$,
\begin{equation}
  \label{eq:ok1+F}
  -i\omega \langle h\rangle + i\vk\bm\cdot
\langle \bm {\dot x}_0 (1+{\cal F})[h]\rangle=0.
\end{equation}
In order to determine the dispersion relation of the mode, in general,
we would need to know $h(\vk,\vp)$, corresponding to this
mode. However, for $\bm k=0$ we already know that the mode is
given by $h_0$. Using Eq.~(\ref{eq:h+F=mu}), we can then write the
frequency of the hydrodynamic mode to leading order in $\vk$ as
\begin{equation}
  \label{eq:2}
  {\o} = {\vk}\bm\cdot 
\frac{\langle  \bm {\dot    x}_0\rangle}{\langle h_0\rangle}  + \OO(k^2)\,.
\end{equation}
Using the definition of the full velocity $\bm{\dot x}_0$ in
Eq.~(\ref{eq:xdot-0}) we see that the first term does not contribute
to the average $\langle  \bm {\dot
    x}_0\rangle$, since it gives an integral of a total derivative,
  and as a result
  \begin{equation}
    \label{eq:x_0}
    \langle  \bm {\dot
    x}_0\rangle 
= \frac1{\chi_0}\int_{\vp} \frac{\pd\npeq}{\pd\mu}
(\dedp_0+\bm B(\dedp_0\bm{\cdot b})) = 
\frac{\bm B}{\chi_0}\int_{\vp} \left(-\frac{\pd\npeq}{\pd\vp}\right)
\bm{\cdot b} = \frac{\bm B}{4\pi^2\chi_0}\,,
  \end{equation}
where we integrated by parts and used $(\pd/\pd\vp)\bm{\cdot
  b}=2\pi\delta^3(\vp)$ and $n_0(\vp=0)=1$. The coefficient $1/4\pi^2$ here
is the same as in the expression for the chiral magnetic effect $\bm J
= 1/(4\pi^2)\mu\bm B$.

The velocity of the hydrodynamic mode (chiral magnetic wave) is given
by
\begin{equation}
  \label{eq:vcmw}
  \vv_\textrm{CMW}=\frac{\pd\omega}{\pd\bm k} 
= \frac{{\bm B}}{4\pi^2\chi_0\langle h_0 \rangle} \, .
\end{equation}
The product ${\chi_0}\langle h_0 \rangle$ is the full susceptibility
$\chi$ of
the Fermi liquid with the interactions. Indeed, by
definition, $\chi$ is the rate of change of the volume density $n_V$
as a function of the chemical potential,
\begin{equation}
  \label{eq:3}
  \chi = \frac{\delta n_V}{\delta \mu}\,,
\end{equation}
where $n_V=\int_{\vp}\sqrt G\, n$.  Thus, using Eqs.~(\ref{eq:h0}),
 (\ref{eq:average}) and the definition of $h_0=\delta h/\delta\mu$,
\begin{equation}
  \label{eq:3}
  \chi = \int_{\vp}\sqrt G\, \frac{\pd\npeq}{\pd\mu}\frac{\delta h}{\delta \mu}
= \chi_0 \langle h_0\rangle \,.
\end{equation}

This shows that the chiral magnetic wave velocity is still given
by the same expression in terms of susceptibility in the cold dense Weyl liquid.
As in the case of finite temperature,
it also has no dependence on the details of dispersion relation
$\e({\vp})$. 
Remarkably,
the same relation between the speed of chiral magnetic wave and the susceptibility
still holds for Fermi liquid even if the susceptibility is modified
due to interaction.

\subsection{Transition of CMW into the zero sound}

In the previous subsection,
we have established the chiral magnetic wave in a cold Fermi liquid system in the hydrodynamic regime $\o, k \ll \tau^{-1}$,
where $\tau$ is the relaxation time due to particle collisions.
In the opposite limit of $\o, k \gg \tau^{-1}$,
the Fermi liquid has another collective excitation -- the 
zero sound.
At low temperature,
a typical relaxation time scales as $\tau\sim{\mu/ T^2}$ due to Pauli blocking effect \cite{Abrikosov:1959,lifshitz1981physical},
and if the temperature is strictly zero, the relaxation time diverges
and one is always in collisionless (zero sound) regime.
We will consider a small,
but finite temperature so that one can discuss the transition
from the chiral magnetic wave at $\omega\ll\tau^{-1}$  to the zero
sound at $\omega\gg\tau^{-1}$.

For this purpose,
let us study the dispersion relation in the kinetic theory within the relaxation time approximation,
\be
\label{eq:RTA-zeroT}
{\cal I}[h] = -\frac{h -\<\, h\,\>}{\tau}\, .
\ee
The general form of the linearized Landau interaction functional ${\cal F}[h]$ to linear order in $B$
 was discussed in Ref.~\cite{Son:2012wh}.
As shown in Ref.~\cite{Baym:1975va},
for massless quark matter weakly interacting via exchange of color gluons in the absence of external magnetic field,
one can approximate
${\cal F}[h]=F_0 \<\, h\,\>$ with a constant $F_0$ to lowest order in
$\a_{s}$. For simplicity, we will assume this form of the Landau functional in the following analysis.

The linearized kinetic equation with these assumptions reads
\be
\label{eq:kinetic-zero1}
\(- i \o + i \dot\vx_0\cdot\vk +\tau^{-1}\)\, h
= \(\,\tau^{-1}  - iF_0\,\dot\vx_0\cdot\vk \,\)\<\, h\,\>\, ,
\ee
where the Lorentz force term disappeared as we assumed, for simplicity, $\vk$ parallel to $\vB$ as before.
The equation for the dispersion relation is 
\be
\label{eq:master1}
\left\langle
\,
\frac{\tau^{-1} - i F_0\,\dot\vx_0\cdot\vk }
{-i\o + i \dot\vx_0\cdot\vk +\tau^{-1}} \,
\right\rangle
= 1 \, .
\ee

It is easy to see that when $\tau^{-1}\to\infty$ (i.e., $\omega\tau\ll1$) we
recover equation~(\ref{eq:ok1+F})  obtained earlier in hydrodynamic
limit.  In the opposite, collisionless limit, $\omega\tau\gg1$, we can
drop $\tau^{-1}$ and the resulting equation is the well-known equation
for the zero sound.

In solving this equation, we consider the temperature low enough that $\partial n_0/\partial \mu$ can be replaced by $\delta(\epsilon_0-\mu)$ up to higher order corrections in $T/\mu$.
Let us first consider the case without magnetic field in the collisionless regime $\omega\gg \tau^{-1}$; in this case it is easy to solve \eqref{eq:master1} to obtain
\be
\o_0(k)=s_0v_F k\,,\qquad v_F\equiv \left|{\partial \epsilon_0\over\partial \vp}\right|_{\epsilon_0=\mu}\,,
\ee
where $s_0$ is a solution of
\be
 \frac{s_0}{2}\log\left(\frac{s_0 -1}{s_0-1}\right) - 1
= F^{-1}_0\,.
\ee
This is the well-known result 
for the zero sound in normal Fermi liquid \cite{Abrikosov:1959}. Since we have shown in small $B$ limit the existence of chiral magnetic wave in the $k\ll\tau^{-1}$ regime, we naturally expect a transition between the chiral magnetic wave and the zero sound as $k$ increases from $k\ll\tau^{-1}$ to $k\gg\tau^{-1}$, which we confirm in our numerical results.
\begin{figure}[t]
  \centering
  \subfigure[]{
          \label{fig:zeroTdisre}
	\includegraphics[width=.45\textwidth]{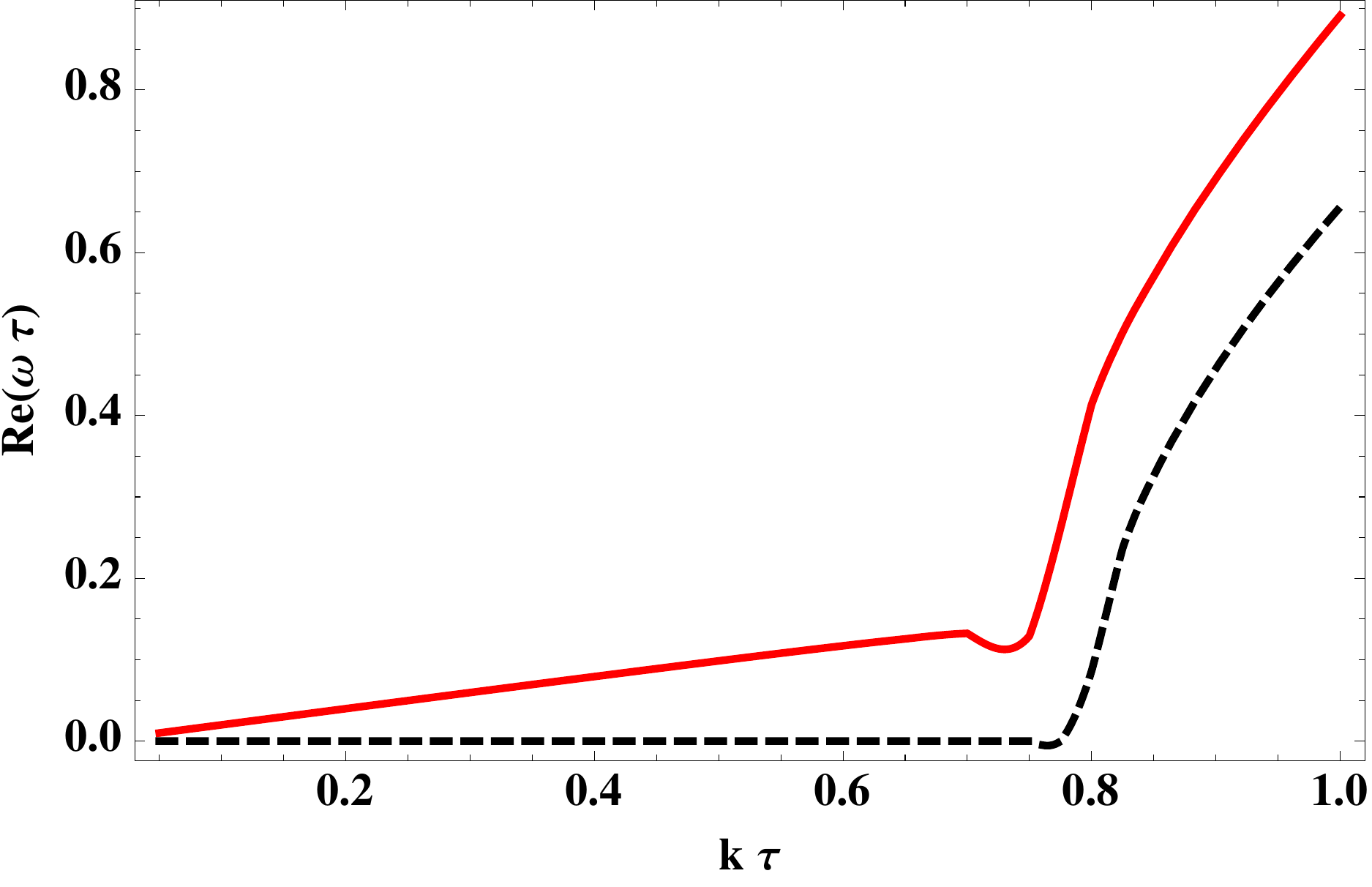}
          }
	\subfigure[]{
          \label{fig:zeroTdisim}
	\includegraphics[width=.45\textwidth]{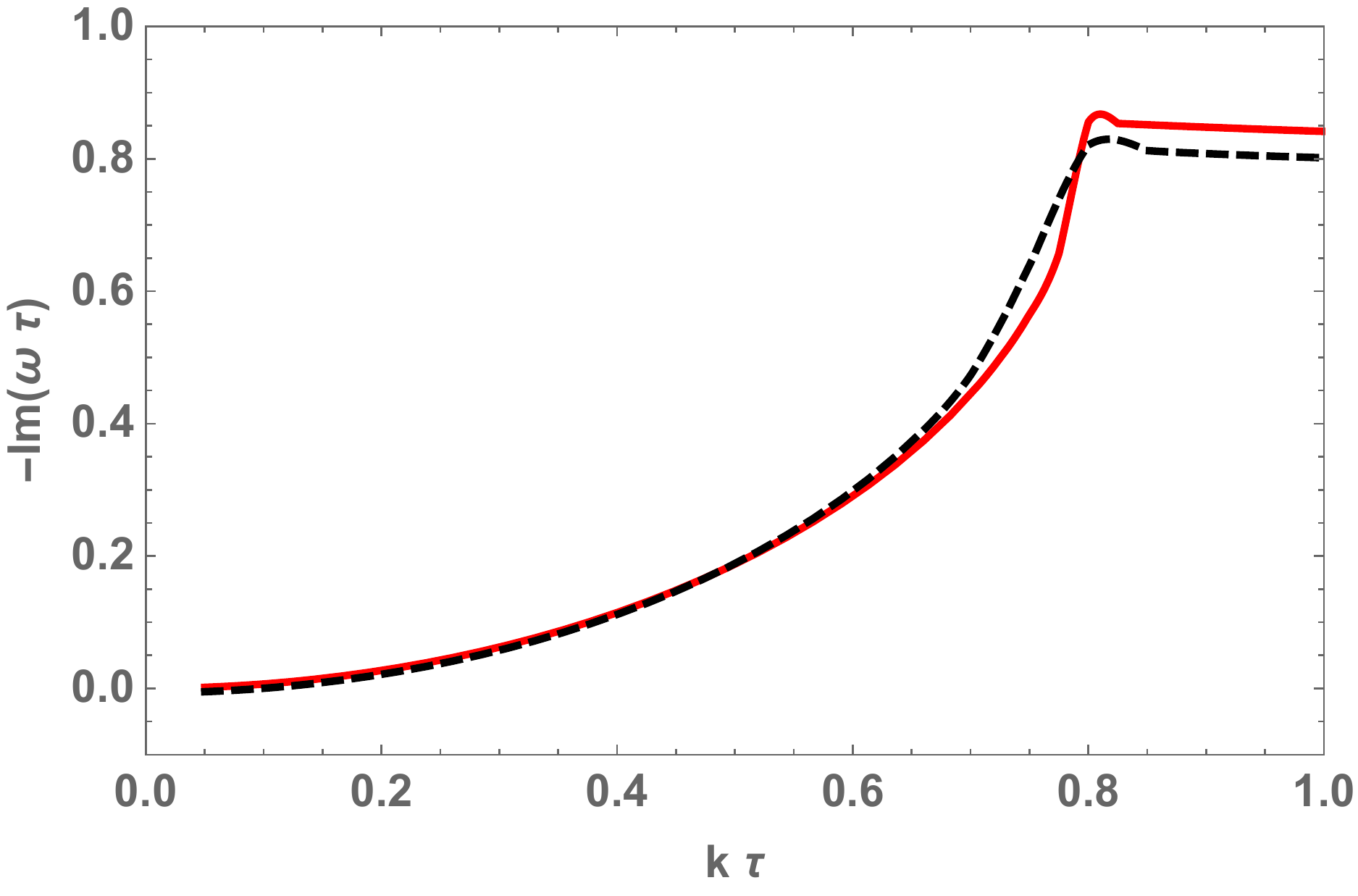}
            }
	\caption{
          \label{fig:zeroTdis}
          The numerical solution of \eq\eqref{eq:master1} for the real and imaginary parts of
          the collective mode frequency $\o(k)$ as functions of the
          wave number (in units of $\tau^{-1}$) in a cold Fermi liquid
          with Landau parameter $F_0=1$ and $v_B\equiv
          {B/4\pi^2\chi}=0.1$. The dashed curves are the same but in
          zero magnetic field.  }
      \end{figure}

In FIG.~\ref{fig:zeroTdis},
we present our numerical results for the dispersion relation $\o(k)$ with $F_0=1$ and $v_B\equiv {B\over 4\pi^2\chi}=0.1$.
Results with a different choice of $(F_0,B)$ are qualitatively similar.
For comparison,
we also plot the dispersion relation without the magnetic field (the dashed curve).
As one can anticipate,  
$\o(k)$ without magnetic field is purely imaginary for sufficiently small $k$ due to diffusion, while 
for a large $k$ one finds the propagating zero sound with
frequency-independent attenuation.
For $B\neq0$, there is a linear dispersion of the CMW for small $k \ll \tau^{-1}$, and
as $k$ increases 
we observe the transition to the zero sound, modified by the
magnetic field.

\section{Summary and conclusion}
\label{sec:conclusion}

In this paper,
we studied collective modes of a weakly coupled plasma of chiral
fermion quasiparticles in the presence of an external magnetic field,
using the framework of the recently developed chiral kinetic theory.
We show the existence of the chiral magnetic wave in both the high
temperature, low density regime, as well as the cold, dense Fermi
liquid case, in a general way without relying on an explicit form of
the collision term. The velocity of the chiral magnetic wave is shown
to be given universally by $\bm v_B=\bm B/(4\pi^2\chi)$ where $\chi$ is the
charge susceptibility, even when the interactions in the Landau Fermi liquid
modify the quasi-particle energies.

In a gas of massless Dirac fermions, we show how inter-chiral
transitions cause chiral magnetic wave to become a
vector-like charge diffusion mode at sufficiently low wave numbers.
In the interacting cold Landau Fermi liquid, we
observe an interesting transition from the chiral magnetic wave to the
zero sound as the wave number increases from the hydrodynamic to the collisionless regime.

One of the important approximations in this work is the decoupling of the charge fluctuations we consider from the energy-momentum fluctuations.
In high temperature, low density regime (discussed in the sections \ref{sec:left} and \ref{sec:sphaleron}),
it can be justified as far as $T\gg \mu$.
In high density, low temperature regime (discussed in the section \ref{sec:fermi}),
this assumption may again be justified for large $N$ theories as the coupling between the charge and the energy-momentum fluctuations is $1/N$ suppressed.
It would be interesting to extend our analysis including the energy-momentum fluctuations to see, for example, the mixing between the first sound and the chiral magnetic wave.
Another interesting direction is to let the electromagnetism be dynamical, 
and study dynamical collective excitations such as plasmons
\cite{Kharzeev:2010gd,Akamatsu:2013pjd} as well as their interplay with the chiral magnetic wave.
We leave these and other extensions of this work to future studies.

\newpage

\vskip 1cm \centerline{\large \bf Acknowledgment} \vskip 0.5cm
We thank Amadeo Jimenez-Alba and Dima Kharzeev for helpful discussions.
This work was supported by the U.S. Department of Energy under Contracts
No.\ DE-FG-88ER40388 and DE-FG0201ER41195.

\bibliography{CMW}

\begin{thebibliography}{47}
\expandafter\ifx\csname natexlab\endcsname\relax\def\natexlab#1{#1}\fi
\expandafter\ifx\csname bibnamefont\endcsname\relax
  \def\bibnamefont#1{#1}\fi
\expandafter\ifx\csname bibfnamefont\endcsname\relax
  \def\bibfnamefont#1{#1}\fi
\expandafter\ifx\csname citenamefont\endcsname\relax
  \def\citenamefont#1{#1}\fi
\expandafter\ifx\csname url\endcsname\relax
  \def\url#1{\texttt{#1}}\fi
\expandafter\ifx\csname urlprefix\endcsname\relax\def\urlprefix{URL }\fi
\providecommand{\bibinfo}[2]{#2}
\providecommand{\eprint}[2][]{\url{#2}}

\bibitem[{\citenamefont{Kharzeev et~al.}(2008)\citenamefont{Kharzeev, McLerran,
  and Warringa}}]{Kharzeev:2007jp}
\bibinfo{author}{\bibfnamefont{D.~E.} \bibnamefont{Kharzeev}},
  \bibinfo{author}{\bibfnamefont{L.~D.} \bibnamefont{McLerran}},
  \bibnamefont{and} \bibinfo{author}{\bibfnamefont{H.~J.}
  \bibnamefont{Warringa}}, \bibinfo{journal}{Nucl.Phys.}
  \textbf{\bibinfo{volume}{A803}}, \bibinfo{pages}{227} (\bibinfo{year}{2008}),
  \eprint{0711.0950}.

\bibitem[{\citenamefont{Fukushima et~al.}(2008)\citenamefont{Fukushima,
  Kharzeev, and Warringa}}]{Fukushima:2008xe}
\bibinfo{author}{\bibfnamefont{K.}~\bibnamefont{Fukushima}},
  \bibinfo{author}{\bibfnamefont{D.~E.} \bibnamefont{Kharzeev}},
  \bibnamefont{and} \bibinfo{author}{\bibfnamefont{H.~J.}
  \bibnamefont{Warringa}}, \bibinfo{journal}{Phys. Rev. D}
  \textbf{\bibinfo{volume}{78}}, \bibinfo{pages}{074033}
  (\bibinfo{year}{2008}), \eprint{0808.3382}.

\bibitem[{\citenamefont{Son and Zhitnitsky}(2004)}]{Son:2004tq}
\bibinfo{author}{\bibfnamefont{D.}~\bibnamefont{Son}} \bibnamefont{and}
  \bibinfo{author}{\bibfnamefont{A.~R.} \bibnamefont{Zhitnitsky}},
  \bibinfo{journal}{Phys.Rev.} \textbf{\bibinfo{volume}{D70}},
  \bibinfo{pages}{074018} (\bibinfo{year}{2004}), \eprint{hep-ph/0405216}.

\bibitem[{\citenamefont{Son and Stephanov}(2008)}]{Son:2007ny}
\bibinfo{author}{\bibfnamefont{D.}~\bibnamefont{Son}} \bibnamefont{and}
  \bibinfo{author}{\bibfnamefont{M.}~\bibnamefont{Stephanov}},
  \bibinfo{journal}{Phys.Rev.} \textbf{\bibinfo{volume}{D77}},
  \bibinfo{pages}{014021} (\bibinfo{year}{2008}), \eprint{0710.1084}.

\bibitem[{\citenamefont{Vilenkin}(1980)}]{Vilenkin:1980fu}
\bibinfo{author}{\bibfnamefont{A.}~\bibnamefont{Vilenkin}},
  \bibinfo{journal}{Phys.Rev.} \textbf{\bibinfo{volume}{D22}},
  \bibinfo{pages}{3080} (\bibinfo{year}{1980}).

\bibitem[{\citenamefont{Kharzeev and Son}(2011)}]{Kharzeev:2010gr}
\bibinfo{author}{\bibfnamefont{D.~E.} \bibnamefont{Kharzeev}} \bibnamefont{and}
  \bibinfo{author}{\bibfnamefont{D.~T.} \bibnamefont{Son}},
  \bibinfo{journal}{Phys.Rev.Lett.} \textbf{\bibinfo{volume}{106}},
  \bibinfo{pages}{062301} (\bibinfo{year}{2011}), \eprint{1010.0038}.

\bibitem[{\citenamefont{Voloshin}(2004)}]{Voloshin:2004vk}
\bibinfo{author}{\bibfnamefont{S.~A.} \bibnamefont{Voloshin}},
  \bibinfo{journal}{Phys.Rev.} \textbf{\bibinfo{volume}{C70}},
  \bibinfo{pages}{057901} (\bibinfo{year}{2004}), \eprint{hep-ph/0406311}.

\bibitem[{\citenamefont{Abelev et~al.}(2009)}]{Abelev:2009ac}
\bibinfo{author}{\bibfnamefont{B.}~\bibnamefont{Abelev}} \bibnamefont{et~al.}
  (\bibinfo{collaboration}{STAR Collaboration}),
  \bibinfo{journal}{Phys.Rev.Lett.} \textbf{\bibinfo{volume}{103}},
  \bibinfo{pages}{251601} (\bibinfo{year}{2009}), \eprint{0909.1739}.

\bibitem[{\citenamefont{Selyuzhenkov}(2012)}]{Selyuzhenkov:2011xq}
\bibinfo{author}{\bibfnamefont{I.}~\bibnamefont{Selyuzhenkov}}
  (\bibinfo{collaboration}{ALICE Collaboration}),
  \bibinfo{journal}{Prog.Theor.Phys.Suppl.} \textbf{\bibinfo{volume}{193}},
  \bibinfo{pages}{153} (\bibinfo{year}{2012}), \eprint{1111.1875}.

\bibitem[{\citenamefont{Adamczyk et~al.}(2014)}]{Adamczyk:2014mzf}
\bibinfo{author}{\bibfnamefont{L.}~\bibnamefont{Adamczyk}} \bibnamefont{et~al.}
  (\bibinfo{collaboration}{STAR Collaboration}),
  \bibinfo{journal}{Phys.Rev.Lett.} \textbf{\bibinfo{volume}{113}},
  \bibinfo{pages}{052302} (\bibinfo{year}{2014}), \eprint{1404.1433}.

\bibitem[{\citenamefont{Son and Surowka}(2009)}]{Son:2009tf}
\bibinfo{author}{\bibfnamefont{D.~T.} \bibnamefont{Son}} \bibnamefont{and}
  \bibinfo{author}{\bibfnamefont{P.}~\bibnamefont{Surowka}},
  \bibinfo{journal}{Phys. Rev. Lett.} \textbf{\bibinfo{volume}{103}},
  \bibinfo{pages}{191601} (\bibinfo{year}{2009}), \eprint{0906.5044}.

\bibitem[{\citenamefont{Buividovich et~al.}(2009)\citenamefont{Buividovich,
  Chernodub, Luschevskaya, and Polikarpov}}]{Buividovich:2009wi}
\bibinfo{author}{\bibfnamefont{P.}~\bibnamefont{Buividovich}},
  \bibinfo{author}{\bibfnamefont{M.}~\bibnamefont{Chernodub}},
  \bibinfo{author}{\bibfnamefont{E.}~\bibnamefont{Luschevskaya}},
  \bibnamefont{and}
  \bibinfo{author}{\bibfnamefont{M.}~\bibnamefont{Polikarpov}},
  \bibinfo{journal}{Phys.Rev.} \textbf{\bibinfo{volume}{D80}},
  \bibinfo{pages}{054503} (\bibinfo{year}{2009}), \eprint{0907.0494}.

\bibitem[{\citenamefont{Abramczyk et~al.}(2009)\citenamefont{Abramczyk, Blum,
  Petropoulos, and Zhou}}]{Abramczyk:2009gb}
\bibinfo{author}{\bibfnamefont{M.}~\bibnamefont{Abramczyk}},
  \bibinfo{author}{\bibfnamefont{T.}~\bibnamefont{Blum}},
  \bibinfo{author}{\bibfnamefont{G.}~\bibnamefont{Petropoulos}},
  \bibnamefont{and} \bibinfo{author}{\bibfnamefont{R.}~\bibnamefont{Zhou}},
  \bibinfo{journal}{PoS} \textbf{\bibinfo{volume}{LAT2009}},
  \bibinfo{pages}{181} (\bibinfo{year}{2009}), \eprint{0911.1348}.

\bibitem[{\citenamefont{Yamamoto}(2011)}]{Yamamoto:2011gk}
\bibinfo{author}{\bibfnamefont{A.}~\bibnamefont{Yamamoto}},
  \bibinfo{journal}{Phys.Rev.Lett.} \textbf{\bibinfo{volume}{107}},
  \bibinfo{pages}{031601} (\bibinfo{year}{2011}), \eprint{1105.0385}.

\bibitem[{\citenamefont{Buividovich}(2014)}]{Buividovich:2013hza}
\bibinfo{author}{\bibfnamefont{P.}~\bibnamefont{Buividovich}},
  \bibinfo{journal}{Nucl.Phys.} \textbf{\bibinfo{volume}{A925}},
  \bibinfo{pages}{218} (\bibinfo{year}{2014}), \eprint{1312.1843}.

\bibitem[{\citenamefont{Bali et~al.}(2014)\citenamefont{Bali, Bruckmann,
  Endršdi, Fodor, Katz et~al.}}]{Bali:2014vja}
\bibinfo{author}{\bibfnamefont{G.}~\bibnamefont{Bali}},
  \bibinfo{author}{\bibfnamefont{F.}~\bibnamefont{Bruckmann}},
  \bibinfo{author}{\bibfnamefont{G.}~\bibnamefont{Endršdi}},
  \bibinfo{author}{\bibfnamefont{Z.}~\bibnamefont{Fodor}},
  \bibinfo{author}{\bibfnamefont{S.}~\bibnamefont{Katz}}, \bibnamefont{et~al.},
  \bibinfo{journal}{JHEP} \textbf{\bibinfo{volume}{1404}}, \bibinfo{pages}{129}
  (\bibinfo{year}{2014}), \eprint{1401.4141}.

\bibitem[{\citenamefont{Sadofyev et~al.}(2011)\citenamefont{Sadofyev,
  Shevchenko, and Zakharov}}]{Sadofyev:2010is}
\bibinfo{author}{\bibfnamefont{A.}~\bibnamefont{Sadofyev}},
  \bibinfo{author}{\bibfnamefont{V.}~\bibnamefont{Shevchenko}},
  \bibnamefont{and} \bibinfo{author}{\bibfnamefont{V.}~\bibnamefont{Zakharov}},
  \bibinfo{journal}{Phys.Rev.} \textbf{\bibinfo{volume}{D83}},
  \bibinfo{pages}{105025} (\bibinfo{year}{2011}), \eprint{1012.1958}.

\bibitem[{\citenamefont{Nair et~al.}(2012)\citenamefont{Nair, Ray, and
  Roy}}]{Nair:2011mk}
\bibinfo{author}{\bibfnamefont{V.}~\bibnamefont{Nair}},
  \bibinfo{author}{\bibfnamefont{R.}~\bibnamefont{Ray}}, \bibnamefont{and}
  \bibinfo{author}{\bibfnamefont{S.}~\bibnamefont{Roy}},
  \bibinfo{journal}{Phys.Rev.} \textbf{\bibinfo{volume}{D86}},
  \bibinfo{pages}{025012} (\bibinfo{year}{2012}), \eprint{1112.4022}.

\bibitem[{\citenamefont{Yee}(2009)}]{Yee:2009vw}
\bibinfo{author}{\bibfnamefont{H.-U.} \bibnamefont{Yee}},
  \bibinfo{journal}{JHEP} \textbf{\bibinfo{volume}{0911}}, \bibinfo{pages}{085}
  (\bibinfo{year}{2009}), \eprint{0908.4189}.

\bibitem[{\citenamefont{Rebhan et~al.}(2010)\citenamefont{Rebhan, Schmitt, and
  Stricker}}]{Rebhan:2009vc}
\bibinfo{author}{\bibfnamefont{A.}~\bibnamefont{Rebhan}},
  \bibinfo{author}{\bibfnamefont{A.}~\bibnamefont{Schmitt}}, \bibnamefont{and}
  \bibinfo{author}{\bibfnamefont{S.~A.} \bibnamefont{Stricker}},
  \bibinfo{journal}{JHEP} \textbf{\bibinfo{volume}{1001}}, \bibinfo{pages}{026}
  (\bibinfo{year}{2010}), \eprint{0909.4782}.

\bibitem[{\citenamefont{Gorsky et~al.}(2011)\citenamefont{Gorsky, Kopnin, and
  Zayakin}}]{Gorsky:2010xu}
\bibinfo{author}{\bibfnamefont{A.}~\bibnamefont{Gorsky}},
  \bibinfo{author}{\bibfnamefont{P.}~\bibnamefont{Kopnin}}, \bibnamefont{and}
  \bibinfo{author}{\bibfnamefont{A.}~\bibnamefont{Zayakin}},
  \bibinfo{journal}{Phys.Rev.} \textbf{\bibinfo{volume}{D83}},
  \bibinfo{pages}{014023} (\bibinfo{year}{2011}), \eprint{1003.2293}.

\bibitem[{\citenamefont{Hoyos et~al.}(2011)\citenamefont{Hoyos, Nishioka, and
  O'Bannon}}]{Hoyos:2011us}
\bibinfo{author}{\bibfnamefont{C.}~\bibnamefont{Hoyos}},
  \bibinfo{author}{\bibfnamefont{T.}~\bibnamefont{Nishioka}}, \bibnamefont{and}
  \bibinfo{author}{\bibfnamefont{A.}~\bibnamefont{O'Bannon}},
  \bibinfo{journal}{JHEP} \textbf{\bibinfo{volume}{1110}}, \bibinfo{pages}{084}
  (\bibinfo{year}{2011}), \eprint{1106.4030}.

\bibitem[{\citenamefont{Kharzeev and Yee}(2011)}]{Kharzeev:2010gd}
\bibinfo{author}{\bibfnamefont{D.~E.} \bibnamefont{Kharzeev}} \bibnamefont{and}
  \bibinfo{author}{\bibfnamefont{H.-U.} \bibnamefont{Yee}},
  \bibinfo{journal}{Phys.Rev.} \textbf{\bibinfo{volume}{D83}},
  \bibinfo{pages}{085007} (\bibinfo{year}{2011}), \eprint{1012.6026}.

\bibitem[{\citenamefont{Newman}(2006)}]{Newman:2005hd}
\bibinfo{author}{\bibfnamefont{G.}~\bibnamefont{Newman}},
  \bibinfo{journal}{JHEP} \textbf{\bibinfo{volume}{0601}}, \bibinfo{pages}{158}
  (\bibinfo{year}{2006}), \eprint{hep-ph/0511236}.

\bibitem[{\citenamefont{Burnier et~al.}(2011)\citenamefont{Burnier, Kharzeev,
  Liao, and Yee}}]{Burnier:2011bf}
\bibinfo{author}{\bibfnamefont{Y.}~\bibnamefont{Burnier}},
  \bibinfo{author}{\bibfnamefont{D.~E.} \bibnamefont{Kharzeev}},
  \bibinfo{author}{\bibfnamefont{J.}~\bibnamefont{Liao}}, \bibnamefont{and}
  \bibinfo{author}{\bibfnamefont{H.-U.} \bibnamefont{Yee}},
  \bibinfo{journal}{Phys.Rev.Lett.} \textbf{\bibinfo{volume}{107}},
  \bibinfo{pages}{052303} (\bibinfo{year}{2011}), \eprint{1103.1307}.

\bibitem[{\citenamefont{Gorbar et~al.}(2011)\citenamefont{Gorbar, Miransky, and
  Shovkovy}}]{Gorbar:2011ya}
\bibinfo{author}{\bibfnamefont{E.}~\bibnamefont{Gorbar}},
  \bibinfo{author}{\bibfnamefont{V.}~\bibnamefont{Miransky}}, \bibnamefont{and}
  \bibinfo{author}{\bibfnamefont{I.}~\bibnamefont{Shovkovy}},
  \bibinfo{journal}{Phys.Rev.} \textbf{\bibinfo{volume}{D83}},
  \bibinfo{pages}{085003} (\bibinfo{year}{2011}), \eprint{1101.4954}.

\bibitem[{\citenamefont{Yee and Yin}(2014)}]{Yee:2013cya}
\bibinfo{author}{\bibfnamefont{H.-U.} \bibnamefont{Yee}} \bibnamefont{and}
  \bibinfo{author}{\bibfnamefont{Y.}~\bibnamefont{Yin}},
  \bibinfo{journal}{Phys.Rev.} \textbf{\bibinfo{volume}{C89}},
  \bibinfo{pages}{044909} (\bibinfo{year}{2014}), \eprint{1311.2574}.

\bibitem[{\citenamefont{Burnier et~al.}(2012)\citenamefont{Burnier, Kharzeev,
  Liao, and Yee}}]{Burnier:2012ae}
\bibinfo{author}{\bibfnamefont{Y.}~\bibnamefont{Burnier}},
  \bibinfo{author}{\bibfnamefont{D.}~\bibnamefont{Kharzeev}},
  \bibinfo{author}{\bibfnamefont{J.}~\bibnamefont{Liao}}, \bibnamefont{and}
  \bibinfo{author}{\bibfnamefont{H.-U.} \bibnamefont{Yee}}
  (\bibinfo{year}{2012}), \eprint{1208.2537}.

\bibitem[{\citenamefont{Wang}(2013)}]{Wang:2012qs}
\bibinfo{author}{\bibfnamefont{G.}~\bibnamefont{Wang}}
  (\bibinfo{collaboration}{STAR Collaboration}),
  \bibinfo{journal}{Nucl.Phys.A904-905} \textbf{\bibinfo{volume}{2013}},
  \bibinfo{pages}{248c} (\bibinfo{year}{2013}), \eprint{1210.5498}.

\bibitem[{\citenamefont{Ke}(2012)}]{Ke:2012qb}
\bibinfo{author}{\bibfnamefont{H.}~\bibnamefont{Ke}}
  (\bibinfo{collaboration}{STAR Collaboration}),
  \bibinfo{journal}{J.Phys.Conf.Ser.} \textbf{\bibinfo{volume}{389}},
  \bibinfo{pages}{012035} (\bibinfo{year}{2012}), \eprint{1211.3216}.

\bibitem[{\citenamefont{Shou}(2014)}]{Shou:2014cua}
\bibinfo{author}{\bibfnamefont{Q.-Y.} \bibnamefont{Shou}}
  (\bibinfo{collaboration}{STAR Collaboration}),
  \bibinfo{journal}{J.Phys.Conf.Ser.} \textbf{\bibinfo{volume}{509}},
  \bibinfo{pages}{012033} (\bibinfo{year}{2014}).

\bibitem[{\citenamefont{Voloshin and Belmont}(2014)}]{Voloshin:2014gja}
\bibinfo{author}{\bibfnamefont{S.~A.} \bibnamefont{Voloshin}} \bibnamefont{and}
  \bibinfo{author}{\bibfnamefont{R.}~\bibnamefont{Belmont}}
  (\bibinfo{year}{2014}), \eprint{1408.0714}.

\bibitem[{\citenamefont{Son and Yamamoto}(2012)}]{Son:2012wh}
\bibinfo{author}{\bibfnamefont{D.~T.} \bibnamefont{Son}} \bibnamefont{and}
  \bibinfo{author}{\bibfnamefont{N.}~\bibnamefont{Yamamoto}},
  \bibinfo{journal}{Phys.Rev.Lett.} \textbf{\bibinfo{volume}{109}},
  \bibinfo{pages}{181602} (\bibinfo{year}{2012}), \eprint{1203.2697}.

\bibitem[{\citenamefont{Stephanov and Yin}(2012)}]{Stephanov:2012ki}
\bibinfo{author}{\bibfnamefont{M.}~\bibnamefont{Stephanov}} \bibnamefont{and}
  \bibinfo{author}{\bibfnamefont{Y.}~\bibnamefont{Yin}},
  \bibinfo{journal}{Phys.Rev.Lett.} \textbf{\bibinfo{volume}{109}},
  \bibinfo{pages}{162001} (\bibinfo{year}{2012}), \eprint{1207.0747}.

\bibitem[{\citenamefont{Son and Yamamoto}(2013)}]{Son:2012zy}
\bibinfo{author}{\bibfnamefont{D.~T.} \bibnamefont{Son}} \bibnamefont{and}
  \bibinfo{author}{\bibfnamefont{N.}~\bibnamefont{Yamamoto}},
  \bibinfo{journal}{Phys.Rev.} \textbf{\bibinfo{volume}{D87}},
  \bibinfo{pages}{085016} (\bibinfo{year}{2013}), \eprint{1210.8158}.

\bibitem[{\citenamefont{Loganayagam and Surowka}(2012)}]{Loganayagam:2012pz}
\bibinfo{author}{\bibfnamefont{R.}~\bibnamefont{Loganayagam}} \bibnamefont{and}
  \bibinfo{author}{\bibfnamefont{P.}~\bibnamefont{Surowka}},
  \bibinfo{journal}{JHEP} \textbf{\bibinfo{volume}{1204}}, \bibinfo{pages}{097}
  (\bibinfo{year}{2012}), \eprint{1201.2812}.

\bibitem[{\citenamefont{Gao et~al.}(2012)\citenamefont{Gao, Liang, Pu, Wang,
  and Wang}}]{Gao:2012ix}
\bibinfo{author}{\bibfnamefont{J.-H.} \bibnamefont{Gao}},
  \bibinfo{author}{\bibfnamefont{Z.-T.} \bibnamefont{Liang}},
  \bibinfo{author}{\bibfnamefont{S.}~\bibnamefont{Pu}},
  \bibinfo{author}{\bibfnamefont{Q.}~\bibnamefont{Wang}}, \bibnamefont{and}
  \bibinfo{author}{\bibfnamefont{X.-N.} \bibnamefont{Wang}},
  \bibinfo{journal}{Phys.Rev.Lett.} \textbf{\bibinfo{volume}{109}},
  \bibinfo{pages}{232301} (\bibinfo{year}{2012}), \eprint{1203.0725}.

\bibitem[{\citenamefont{Chen et~al.}(2014)\citenamefont{Chen, Son, Stephanov,
  Yee, and Yin}}]{Chen:2014cla}
\bibinfo{author}{\bibfnamefont{J.-Y.} \bibnamefont{Chen}},
  \bibinfo{author}{\bibfnamefont{D.~T.} \bibnamefont{Son}},
  \bibinfo{author}{\bibfnamefont{M.~A.} \bibnamefont{Stephanov}},
  \bibinfo{author}{\bibfnamefont{H.-U.} \bibnamefont{Yee}}, \bibnamefont{and}
  \bibinfo{author}{\bibfnamefont{Y.}~\bibnamefont{Yin}},
  \bibinfo{journal}{Phys.Rev.Lett.} \textbf{\bibinfo{volume}{113}},
  \bibinfo{pages}{182302} (\bibinfo{year}{2014}), \eprint{1404.5963}.

\bibitem[{\citenamefont{Jimenez-Alba et~al.}(2014)\citenamefont{Jimenez-Alba,
  Landsteiner, and Melgar}}]{Jimenez-Alba:2014iia}
\bibinfo{author}{\bibfnamefont{A.}~\bibnamefont{Jimenez-Alba}},
  \bibinfo{author}{\bibfnamefont{K.}~\bibnamefont{Landsteiner}},
  \bibnamefont{and} \bibinfo{author}{\bibfnamefont{L.}~\bibnamefont{Melgar}},
  \bibinfo{journal}{Phys.Rev.} \textbf{\bibinfo{volume}{D90}},
  \bibinfo{pages}{126004} (\bibinfo{year}{2014}), \eprint{1407.8162}.

\bibitem[{\citenamefont{Gorsky and Zayakin}(2013)}]{Gorsky:2012gi}
\bibinfo{author}{\bibfnamefont{A.}~\bibnamefont{Gorsky}} \bibnamefont{and}
  \bibinfo{author}{\bibfnamefont{A.}~\bibnamefont{Zayakin}},
  \bibinfo{journal}{JHEP} \textbf{\bibinfo{volume}{1302}}, \bibinfo{pages}{124}
  (\bibinfo{year}{2013}), \eprint{1206.4725}.

\bibitem[{\citenamefont{Satow and Yee}(2014)}]{Satow:2014lva}
\bibinfo{author}{\bibfnamefont{D.}~\bibnamefont{Satow}} \bibnamefont{and}
  \bibinfo{author}{\bibfnamefont{H.-U.} \bibnamefont{Yee}},
  \bibinfo{journal}{Phys.Rev.} \textbf{\bibinfo{volume}{D90}},
  \bibinfo{pages}{014027} (\bibinfo{year}{2014}), \eprint{1406.1150}.

\bibitem[{\citenamefont{Son and Spivak}(2013)}]{Son:2012bg}
\bibinfo{author}{\bibfnamefont{D.}~\bibnamefont{Son}} \bibnamefont{and}
  \bibinfo{author}{\bibfnamefont{B.}~\bibnamefont{Spivak}},
  \bibinfo{journal}{Phys.Rev.} \textbf{\bibinfo{volume}{B88}},
  \bibinfo{pages}{104412} (\bibinfo{year}{2013}), \eprint{1206.1627}.

\bibitem[{\citenamefont{Landsteiner et~al.}(2014)\citenamefont{Landsteiner,
  Liu, and Sun}}]{Landsteiner:2014vua}
\bibinfo{author}{\bibfnamefont{K.}~\bibnamefont{Landsteiner}},
  \bibinfo{author}{\bibfnamefont{Y.}~\bibnamefont{Liu}}, \bibnamefont{and}
  \bibinfo{author}{\bibfnamefont{Y.-W.} \bibnamefont{Sun}}
  (\bibinfo{year}{2014}), \eprint{1410.6399}.

\bibitem[{\citenamefont{Abrikosov and Khalatnikov}(1959)}]{Abrikosov:1959}
\bibinfo{author}{\bibfnamefont{A.~A.} \bibnamefont{Abrikosov}}
  \bibnamefont{and} \bibinfo{author}{\bibfnamefont{I.~M.}
  \bibnamefont{Khalatnikov}}, \bibinfo{journal}{Reports on Progress in Physics}
  \textbf{\bibinfo{volume}{22}}, \bibinfo{pages}{329} (\bibinfo{year}{1959}),
  \urlprefix\url{http://stacks.iop.org/0034-4885/22/i=1/a=310}.

\bibitem[{\citenamefont{Lifshitz et~al.}(1981)\citenamefont{Lifshitz,
  Pitaevskii, and Landau}}]{lifshitz1981physical}
\bibinfo{author}{\bibfnamefont{E.}~\bibnamefont{Lifshitz}},
  \bibinfo{author}{\bibfnamefont{L.}~\bibnamefont{Pitaevskii}},
  \bibnamefont{and} \bibinfo{author}{\bibfnamefont{L.}~\bibnamefont{Landau}},
  \emph{\bibinfo{title}{Physical kinetics}}, vol.~\bibinfo{volume}{60}
  (\bibinfo{publisher}{Pergamon press Oxford}, \bibinfo{year}{1981}).

\bibitem[{\citenamefont{Baym and Chin}(1976)}]{Baym:1975va}
\bibinfo{author}{\bibfnamefont{G.}~\bibnamefont{Baym}} \bibnamefont{and}
  \bibinfo{author}{\bibfnamefont{S.~A.} \bibnamefont{Chin}},
  \bibinfo{journal}{Nucl.Phys.} \textbf{\bibinfo{volume}{A262}},
  \bibinfo{pages}{527} (\bibinfo{year}{1976}).

\bibitem[{\citenamefont{Akamatsu and Yamamoto}(2013)}]{Akamatsu:2013pjd}
\bibinfo{author}{\bibfnamefont{Y.}~\bibnamefont{Akamatsu}} \bibnamefont{and}
  \bibinfo{author}{\bibfnamefont{N.}~\bibnamefont{Yamamoto}},
  \bibinfo{journal}{Phys.Rev.Lett.} \textbf{\bibinfo{volume}{111}},
  \bibinfo{pages}{052002} (\bibinfo{year}{2013}), \eprint{1302.2125}.

\end{thebibliography}

\end{document}